\documentclass[twocolumn]{aastex631}
\shorttitle{Star Formation in the Extreme Outer Galaxy}
\shortauthors{Izumi et al.}
\graphicspath{{./}{figures/}}

\begin{document}

\title{Overview Results of JWST Observations of Star-Forming Clusters in the Extreme Outer Galaxy}



\author[0000-0003-1604-9127]{Natsuko Izumi}
\affiliation{Institute of Astronomy and Astrophysics, Academia Sinica, No. 1, Section 4, Roosevelt Road, Taipei 10617, Taiwan}
\affiliation{Faculty of Engineering, Gifu University, 1-1 Yanagido, Gifu 501-1193, Japan}
\affiliation{National Astronomical Observatory of Japan, National Institutes of Natural Sciences, 2-21-1 Osawa, Mitaka, Tokyo 181-8588, Japan}

\author[0000-0001-5644-8830]{Michael E. Ressler}
\affiliation{Jet Propulsion Laboratory, California Institute of Technology \\
4800 Oak Grove Drive, Pasadena, CA 91109, USA}

\author[0000-0003-0778-0321]{Ryan M. Lau}
\affiliation{NSF’s NOIRLab, 950 N. Cherry Avenue, Tucson, AZ 85719, USA}
\affiliation{Institute of Space \& Astronautical Science, Japan Aerospace Exploration Agency, 3-1-1 Yoshinodai, Chuo-ku, Sagamihara, Kanagawa 252-5210, Japan}

\author[0000-0003-2777-5861]{Patrick M. Koch}
\affiliation{Institute of Astronomy and Astrophysics, Academia Sinica, No. 1, Section 4, Roosevelt Road, Taipei 10617, Taiwan}

\author[0000-0003-0769-8627]{Masao Saito}
\affiliation{National Astronomical Observatory of Japan, National Institutes of Natural Sciences, 2-21-1 Osawa, Mitaka, Tokyo 181-8588, Japan}
\affiliation{Department of Astronomical Science, School of Physical Science, SOKENDAI (The Graduate University for Advanced Studies), 2-21-1 Osawa, Mitaka, Tokyo 181-8588,
Japan}

\author[0000-0003-4578-2619]{Naoto Kobayashi}
\affiliation{Institute of Astronomy, School of Science, University of Tokyo, 2-21-1 Osawa, Mitaka, Tokyo 181-0015, Japan}
\affiliation{Kiso Observatory, Institute of Astronomy, School of Science, University of Tokyo, 10762-30 Mitake, Kiso-machi, Kiso-gun, Nagano 397-0101, Japan}

\author[0000-0003-3579-7454]{Chikako Yasui}
\affiliation{National Astronomical Observatory of Japan, National Institutes of Natural Sciences, 2-21-1 Osawa, Mitaka, Tokyo 181-8588, Japan}

\begin{abstract}
The extreme outer Galaxy (EOG), which we define as the region of the Milky Way with a galactocentric radius of more than 18 kpc,
provides an excellent opportunity to study star formation in an environment significantly different from that in the solar neighborhood because of its lower metallicity and lower gas density. 
We carried out near- and mid-infrared (NIR and MIR) imaging observations toward two star-forming clusters located in the EOG using JWST NIRCam and MIRI with nine filters:
F115W, F150W, F200W, F350W, F405N, F444W, F770W, F1280W, and F2100W.
In this paper, we present an overview of the observations, data reduction, and initial results.
The NIR sensitivity is approximately 10--80 times better than our previous observation with the Subaru 8.2 m telescope.
Accordingly, the mass detection limit reaches to about 0.01--0.05 $M_\odot$,
which is about 10 times better than the previous observations.
At MIR wavelengths, the high sensitivity and resolution data enable us to resolve individual young stellar objects in such a distant region for the first time.
The mass detection limit at MIR F770W filter reaches about 0.1--0.3 $M_\odot$.
With these new observations, we have identified components of the clusters that previous surveys did not detect,
including class 0 candidates, outflow/jet components, and distinctive nebular structures.
These data will enable us to investigate the properties of star formation in the EOG at the same depth of detail as previous observations of star formation in the solar neighborhood.
\end{abstract}



\section{Introduction} \label{sec:intro}
The extreme outer Galaxy (EOG), which is defined as the region with a galactocentric radius ($R_{\rm G}$) of more than 18 kpc \citep[e.g.,][]{Kobayashi2008,Yasui2008},
provides an excellent opportunity to study star formation in an environment significantly different from the solar neighborhood.
For example, the gas density and metallicity in the EOG are significantly lower than in the solar neighborhood \citep[e.g.,][]{Wolfire2003,Heyer2015,Nakanishi2016,Fernandez2017,Wenger2019,Arellano2020}.
The interstellar medium (ISM) is dominated by \ion{H}{1}, so that H$_2$ fractions are extremely small.
From the radial profile of gas surface densities in our Galaxy \citep[e.g.,][]{Heyer2015,Nakanishi2016}, the \ion{H}{1} and H$_2$ gas surface densities at $R_{\rm G}\sim 18$~kpc are about one-tenth and less than one-tenth of those in the solar neighborhood, respectively.
From the radial profile of the metallicity \citep[e.g.,][]{Fernandez2017,Wenger2019,Arellano2020}, the metallicity at $R_{\rm G} \sim18$ kpc is somewhat less than one third of that in the solar neighborhood.
Furthermore, it is also known that the EOG has less intense ultraviolet fields and a smaller cosmic-ray flux \citep[e.g.,][]{Bloemen1984}.
Such environments may have characteristics similar to those of dwarf galaxies, and to our Galaxy during the early phases of its formation, in particular, in the thick disk formation phase \citep[e.g.,][]{Ferguson1998aj,Buser2000,Kobayashi2008,Xiang2022}.
Therefore, we can directly observe the galaxy formation processes in unprecedented detail at a much closer distance by observing the outskirts of our galaxy rather than other more distant galaxies.

Previous studies investigated the properties of young star-forming clusters in the EOG mainly with large ground-based telescopes \citep[e.g.,][]{Brand2007,Yasui2008,Yasui2009,Yasui2010,Izumi2014,Shimonishi2021}.
For example, based on near-infrared (NIR) imaging observations with the Subaru 8.2~m telescope, \citet{Yasui2008} reported that
the initial mass function (IMF) of star-forming clusters in the EOG
(down to the detection limit of about 0.1 $M_\odot$) is not significantly different from the typical IMFs of nearby clusters.
However, \citet{Yasui2010} discovered that the lifetime of circumstellar disks in the EOG is shorter than that in nearby clusters, also with NIR data from the Subaru telescope.
These short disk lifetimes in low-metallicity environments might be the basis of the planet-metallicity correlation, i.e., the fraction of giant-planet-hosting stars increases with metallicity \citep[e.g.,][]{Fischer2005,Johnson2010}.
However, the specific physical processes that lead to this shortened lifetime are not yet understood.

Statistical studies of star-formation properties in the EOG are also performed using large mid-infrared (MIR) and far-infrared (FIR) space-based observatories \citep[e.g.,][]{Anderson2014,Izumi2017,Izumi2022,Elia2022}.
For example, \citet{Elia2022} reported the radial profile of star-formation rate (SFR) in our 
Galaxy with data from the Herschel Infrared Galactic Plane Survey \citep[Hi-GAL;][]{Molinari2010}.
The profile shows that the SFR in the EOG is much smaller (less than one-tenth) than that in the solar neighborhood.
\citet{Izumi2022} investigate the star-formation efficiency (SFE) converting from H$_2$ gas mass to stellar mass
based on data from the Wide-field Infrared Survey Explorer (WISE) MIR all-sky survey \citep{Wright2010}
and suggest that the SFE in the EOG is not significantly different from that in the solar neighborhood,
despite the SFR being so much lower and in a different environment.

To get a further understanding of the properties of star formation in the EOG,
we performed JWST NIR and MIR imaging observations of two star-forming molecular clouds, Digel Clouds 1 and 2 (hereafter Clouds 1 and 2; their locations on the sky are shown in Figure \ref{target2}).
Utilizing the high sensitivity of JWST, we are able to detect young stellar objects (YSOs) with a detection limit of less than 0.1 $M_\odot$,
which roughly corresponds to the typical peak value of the IMF of the clusters in the solar neighborhood \citep[e.g.,][]{Bastian2010}.
Therefore, we will be able to investigate the peak value of the IMF and the shape of the low-mass ($<$ 0.1 $M_\odot$) end of the IMF in the EOG for the first time.
Furthermore, the high spatial resolution of JWST will enable us to make the first spatially resolved observations of each YSO in the EOG even at the longer wavelengths ($\sim$ 21 $\micron$).
The NIR ($\lambda$ $\sim$ 0.7--2.5 $\micron$) emission traces the warm and hot dust (typically a few 100~K to 1500~K) in the inner part of the circumstellar disk
\citep[within a stellocentric distance of $\sim$ 0.1--1~AU; e.g.,][]{Dullemond2010}, while the MIR ($\lambda$ $\sim$ 3--25~$\micron$) emission can trace a much larger distance in the disk (stellocentric distance of $\sim$ 10 AU).
Therefore, the combination of NIR and MIR data with JWST is crucial for investigating the evolutionary state (including disk lifetime) for each YSO in the EOG \citep[e.g.,][]{Robitaille2006,Koenig2014}.

In this paper, we present a description of the new JWST observations and their first results.
First, we offer an overview of our targets in Section \ref{sec:tar}.
Sections \ref{sec:obs} and \ref{sec:data_red} describe the basic observational setup and data reduction, respectively.
Section \ref{sec:res} presents the observation results.
In Section \ref{sec:dis}, we discuss the properties of structures newly detected by the observations.
Our main conclusions are summarized in Section \ref{sec:sum}.

\section{Overview of the targets} \label{sec:tar}
Clouds 1 and 2 were discovered by \citet{Digel1994} using CO observations with the Harvard-Smithsonian  Center for Astrophysics 1.2~m telescope of distant \ion{H}{1} peaks identified in the Maryland-Green Bank survey \citep{Westerhout1982}.
The detailed structure of those molecular clouds was investigated by CO observations with the Nobeyama Radio Observatory (NRO) 45~m telescope with $\sim$ 1 pc resolution
\citep[][]{Izumi2014,Izumi2017}.
Figure \ref{FoV} shows the CO distributions of Clouds 1 and 2 (Cloud 1: top right panel, Cloud 2: bottom right panel) with their \ion{H}{1} distributions around them
(Cloud 1: top left panel, Cloud 2: bottom left panel).

\citet{Izumi2014} discovered two young embedded clusters, Cloud 1a and Cloud 1b (top right panel of Figure \ref{FoV}), using deep NIR imaging observations with the Subaru 8.2~m telescope aimed at the two CO peaks of Cloud 1.
\citet{Kobayashi2000} found a number of associated infrared sources to confirm the star-forming activity in Cloud 2 by NIR imaging with the QUIST 0.6~m telescope and NIR spectroscopy with the UKIRT 3.8 m telescope.
Extensive wide-field NIR images obtained with the University of Hawaii 2.2~m telescope, encompassing the entirety of Cloud 2, 
led to the identification of two young embedded star clusters located at the two CO peaks of Cloud 2,
namely the Cloud 2N and Cloud 2S clusters (bottom right panel of Figure \ref{FoV}), as reported by \citet{Kobayashi2008}. 
Further insights into the properties of the Cloud 2N and 2S clusters were discussed by \citet{Yasui2009} based on deeper NIR imaging obtained with the Subaru 8.2~m telescope, which covered these clusters. 
The basic properties of those clusters and molecular clouds are summarized in Table \ref{prop-tar}.
In terms of longer IR wavelengths, Spitzer \citep{Werner2004} Infrared Array Camera \citep[IRAC;][]{Fazio2004} data are available for Cloud 2 \citep[Proposal ID: 249;][]{https://doi.org/10.26131/irsa3}, and WISE MIR all-sky survey data \citep{https://doi.org/10.26131/irsa1} are available for both targets.

Cloud 1 is a candidate for the most distantly known star-forming region from the center of our galaxy with $R_{\rm G} \sim 22$~kpc \citep{Digel1994,Izumi2014}.
It is known to be associated with a very large \ion{H}{1} shell \citep[size: $\sim$ 7$^\circ$ $\times$ 3$^\circ$.5, corresponding to $\sim$ 2 kpc $\times$ 1 kpc; top left panel of Figure \ref{FoV};][]{Morras1998,Izumi2014}.
This shell is thought to be produced by the collision between the high-velocity cloud Complex H \citep[HVC131$+$1-200;][]{Hulsbosch1971, Dieter1971} and the Galactic disk \citep{Morras1998}.
\citet{Izumi2014} reported that Cloud 1 is also located very close to the \ion{H}{1} peak of Complex H on the sky with a separation of only $\sim$ 0$^\circ$.5,
and suggested that the impact of Complex H onto the Galactic disk could form Cloud 1 and trigger its star formation activities.
Theoretical and observational studies \citep{Inoue2018,Tokuda2019} suggest that star-forming regions triggered by the shock compression in colliding molecular flows show sequential star-formation associated with the filament structures \citep[Figure 6 in][]{Tokuda2019}.
Note that, in theory, we could expect the same structures even with \ion{H}{1} gas collisions because the basic equations that govern the molecular cloud collisions and \ion{H}{1} cloud collisions are essentially very similar (private communication with T. Inoue). 
Therefore, investigating the evolutionary processes of YSOs in Cloud 1 is crucial for understanding the triggered star-formation process in this region.

Compared with Cloud 1, Cloud 2 shows a large molecular cloud mass and a large number of cluster members (see Table \ref{prop-tar}).
The IMF and lifetime of circumstellar disks were investigated with NIR imaging data from the Subaru 8.2~m telescope by \citet{Yasui2008} and \citet{Yasui2010}, respectively, as mentioned in Section \ref{sec:intro}.
Therefore, Cloud 2 is also a suitable target for a detailed examination of IMF and disk lifetime with JWST.
Cloud 2 is known to be associated with the expanding \ion{H}{1} super shell GSH 138-01-94
\citep[radius: $\sim$ 130 pc, expansion velocity: 10 km s$^{-1}$; bottom left panel of Figure \ref{FoV};][]{Stil2001}, which is driven by a supernova.
Consequently, star formation in Cloud 2 is thought to be triggered by compression from the supernova remnant (SNR) \citep{Kobayashi2008}.

Near Cloud 2N, there exists an isolated early B-type star named MR~1, alongside an \ion{H}{2} region \citep[][; red star mark in the bottom-right panel of Figure \ref{FoV}]{Smartt1996,deGeus1993,Kobayashi2000,Kobayashi2008}.
These features are thought to have been triggered by the SNR prior to the formation of Cloud 2N, and it is hypothesized that the stellar winds emitted from MR~1 might have influenced the formation of Cloud 2N \citep{Kobayashi2008}.

Furthermore, the estimated stellar density suggests that different types of star formation are occurring:
an ``isolated-type" star formation, akin to the Taurus dark cloud \citep[e.g.,][]{Lada1993}, is ongoing in Cloud 2N, 
while ``cluster-type" star formation, resembling the $\rho$~Oph star-forming region \citep[e.g.,][]{Allen2007},
is ongoing in Cloud 2S \citep{Kobayashi2008, Yasui2008}.
Therefore, by including also Cloud 1, we are able to investigate various modes of star formation in the extreme environments of the EOG with JWST.

\section{Observations} \label{sec:obs}
\begin{figure*}
\epsscale{1.1}
\plotone{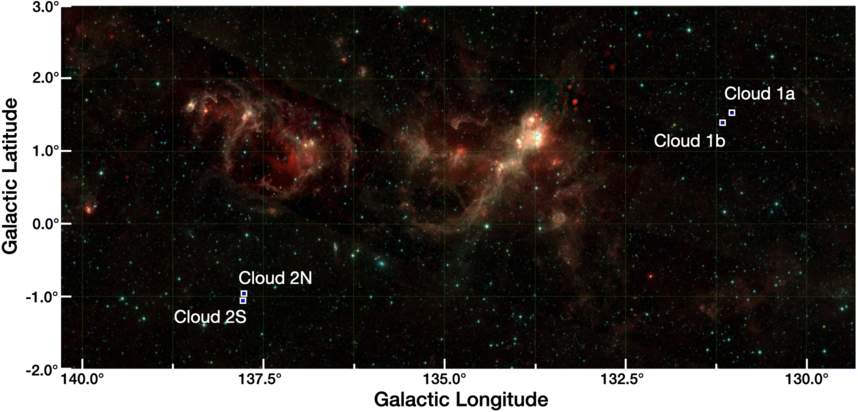}
\caption{
Larger surroundings of our targets:
Digel Clouds 1 and 2 in galactic coordinates.
The background is the WISE w1,w2, and w4 filters pseudocolor image.
This image was made by Astronomers Proposal Tool Aladin Viewer \citep{Bonnarel2000,Boch2014}.
}
\label{target2} 
\end{figure*}

\begin{figure*}
\epsscale{2.0}
\gridline{\fig{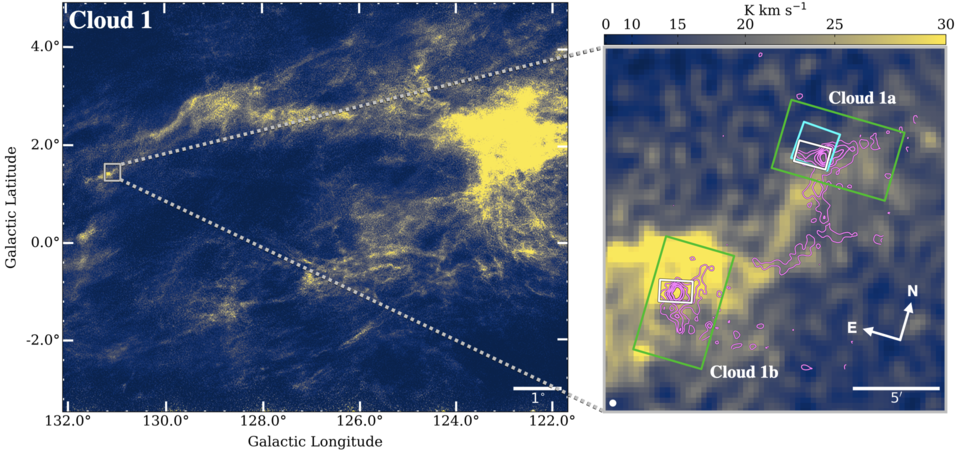}{1.0\textwidth}{}} \vspace{-9mm}
\gridline{\fig{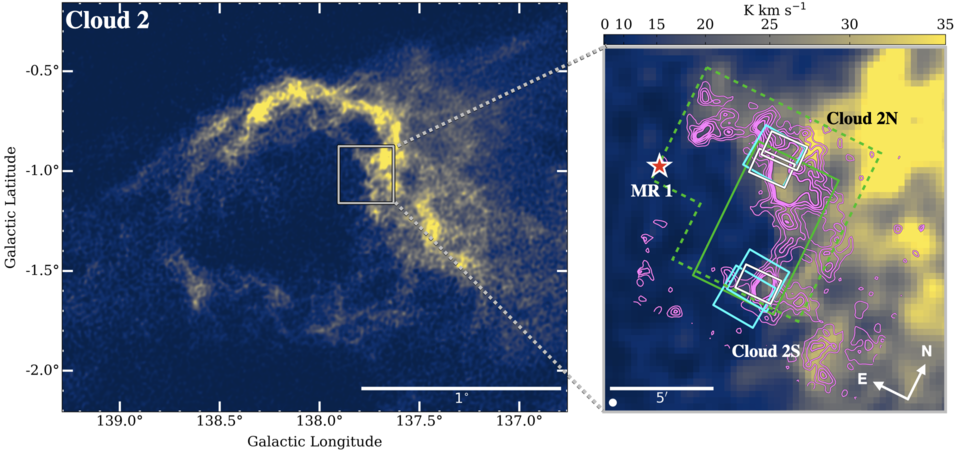}{1.0\textwidth}{} \vspace{-9mm}} 
\caption{
Left: 
\ion{H}{1} distribution around Clouds 1 (top) and 2 (bottom) from Canadian Galactic Plane Survey (CGPS) data \citep[][Cloud 1: $v_{LSR}$ = -104.5 -- -99.6 km s$^{-1}$,
Cloud 2: $v_{LSR}$ = -103.7 -- -96.3 km s$^{-1}$]{Taylor2003}.
The gray squares indicate the locations of Clouds 1 and 2.
Right: 
Zoomed-in images (top: Cloud 1, bottom: Cloud 2) of the gray squares in the left panels.
The magenta contours in the top panel show the integrated $^{12}$CO(1-0) map of Digel Cloud 1 from our NRO 45m data \citep[($V_{\rm LSR}$ = -105.4 -- -98.9 km s$^{-1}$;][]{Izumi2014,Izumi2017}.
Contour levels are 3$\sigma$, 5$\sigma$, 7$\sigma$, 9$\sigma$, and 11$\sigma$, where 1$\sigma$ = 0.92 K km s$^{-1}$. 
The magenta contours in the bottom panel show the integrated $^{12}$CO(1-0) map of Digel Cloud 2 from our NRO-45m data
\citep[$V_{\rm LSR}$ = -106.1 -- -99.1 km s$^{-1}$;][]{Izumi2017}, with contour levels at 3$\sigma$, 5$\sigma$, 7$\sigma$, 9$\sigma$, and 11$\sigma$, where  1$\sigma$ = 1.35 K km s$^{-1}$. 
Contour widths increase with contour levels.
The white and cyan rectangles in both panels represent the field of view (FoV) of the MIRI and NIRCam observations, respectively.
The red star in the bottom right panel marks the location of MR 1 \citep[R.A.:02h49m22.35, decl:+58d26m44.6s;][]{Kobayashi2000}.
The green rectangles in both panels are the FoV of the previous Subaru observations.
The green dotted polygon in the bottom panel is the FoV of the previous observation with the University of Hawaii telescope.
The white filled circle in both panels shows the spatial resolution of the NRO-45m telescope ($\sim$ 17$^{\prime \prime}$).
}
\label{FoV} 
\end{figure*}

Imaging observations were conducted using the JWST MIRI and NIRCam instruments as part of the Cycle~1 
Guaranteed Time Observation (GTO) Program (Program ID: 1237, PI: M.~Ressler). 
Our targets were the two main clusters within Clouds 1 and 2  (white squares in Figure \ref{target2}):
Cloud 1a and 1b in Cloud 1, and Cloud 2N and Cloud 2S in Cloud 2 (white and cyan squares in the right panels of Figure \ref{FoV}).
The observations toward Clouds 1 and 2 took place on 2023 January 10 and 17 (UT), respectively.
The filters employed for NIRCam were F115W, F150W, and F200W in the short channel, and F356W, F444W, 
and F405N in the long channel. 
For MIRI, the filters utilized were F770W, F1280W, and F2100W.

Owing to the coincidence of the spatial distance between Cloud 2N and 2S matching the field-of-view (FoV) separation between NIRCam and MIRI, we were able to employ coordinated parallel observations, where 
NIRCam observations were directed toward Cloud 2S and MIRI observations toward Cloud 2N (using the Prime Template: NIRCam). 
This approach allowed us to optimize observation time since we were allocation-time limited.
The observations of 2N with MIRI and 2S with NIRCam were then done individually, as were all Cloud 1 observations.
To ensure consistency in both luminosity and backgrounds in all filters, we requested grouped observations for each target within a 24 hr period.
The aperture Position Angle (PA) Range was set between 1$^\circ$-20$^\circ$ to encompass all cluster members within the FoV.

To address the gap between detectors in the NIRCam short-wavelength channel, we positioned the cluster within one of the detectors to prevent cluster members from falling into the gap. Additionally, for NIRCam observations, we implemented subpixel dithers to optimally sample the point-spread function (PSF) and enhance the spatial resolution of the stacked images.

Comprehensive details of the observation parameters, including the dither pattern, mosaic number, readout pattern, and exposure time, are provided in Table \ref{obs-para} in Appendix \ref{sec:a-1}

\section{Data Reduction and Photometry} \label{sec:data_red}
Data calibration was performed with the JWST Science Calibration pipeline version number 1.9.5 \citep{Bushouse2023_ver195}, 
the Calibration Reference Data System (CRDS) version 11.16.20,
and CRDC context jwst\_1093.pmap.
We generally used the default values of the parameters in the pipelines except for several custom parameters in the ``jump'' step of the Detector1 pipeline to more aggressively reduce the impact of cosmic-ray showers and ``abs\_refcat'' in the ``tweakreg'' step of the Image3 pipeline, which selects an astrometric reference catalog to better register the images.
We utilized `GAIADR2' to query a GAIA-based astrometric catalog web service for all astrometrically measured sources to improve the registration of all the input images when creating the combined FoV images.

We also inserted a custom background subtraction step between the Image2 and Image3 pipelines. Since there is a significant amount of nebulosity in the images, this custom step uses only portions of the images where there is little nebulosity to establish the background level. The Image3 pipeline then uses these background-subtracted images to build the image mosaics.

We chose to utilize PSF photometry to measure source fluxes due to the high
source-surface density present in all filters.  
For NIRCam, we followed the empirical approach developed by \citet{Anderson2000} for the Hubble Space Telescope Wide-Field Planetary Camera 2 (WFPC2) to obtain a model of the effective pixel-convolved PSF (ePSF) using the affiliated package of Astropy: photoutils \citep{Bradley2019, Bradley2022}\footnote{https://photutils.readthedocs.io/en/stable/}.
To derive the model, we selected stars that were bright but not saturated, that were not close to the edge of the frame, and that did not have any nearby stars within 10 pixels.
However, for MIRI we used PSF models generated from WebbPSF \citep{Perrin2014} since we could not detect a sufficient number of stars to satisfy the criteria mentioned above in these data.
Photometric errors are calculated from the fitting error and
the resampled total-error (ERR) array.
All photometric analyses were conducted on
the Stage3 images that are created after running the Image3 pipeline
to ensure the best sensitivity.
The resulting 10$\sigma$ limiting magnitudes and the FWHM of the PSF across all filters are summarized in Tables \ref{sensitivity} and \ref{fwhm}, respectively.
The corresponding 10$\sigma$ limiting magnitudes in the  $J$ (1.25 \micron), $H$ (1.65 \micron), and $K_{\rm S}$ (2.15 \micron) bands for Cloud 1 obtained from our previous observations with the Subaru telescope are 21.0, 20.5, and 19.5 mag, respectively \citep{Izumi2014}, while those for Cloud 2  are 22.2-22.3, 21.3--21.7, and 21.0--21.2 mag, respectively \citep{Yasui2009}.
Therefore, the NIR sensitivity of our new observations with JWST is about 10--80 times better than our previous observations with Subaru.

\section{Results}\label{sec:res}
\subsection{Entire Images}\label{sec:res-1}
\begin{figure*}
\epsscale{2.0}
\gridline{\fig{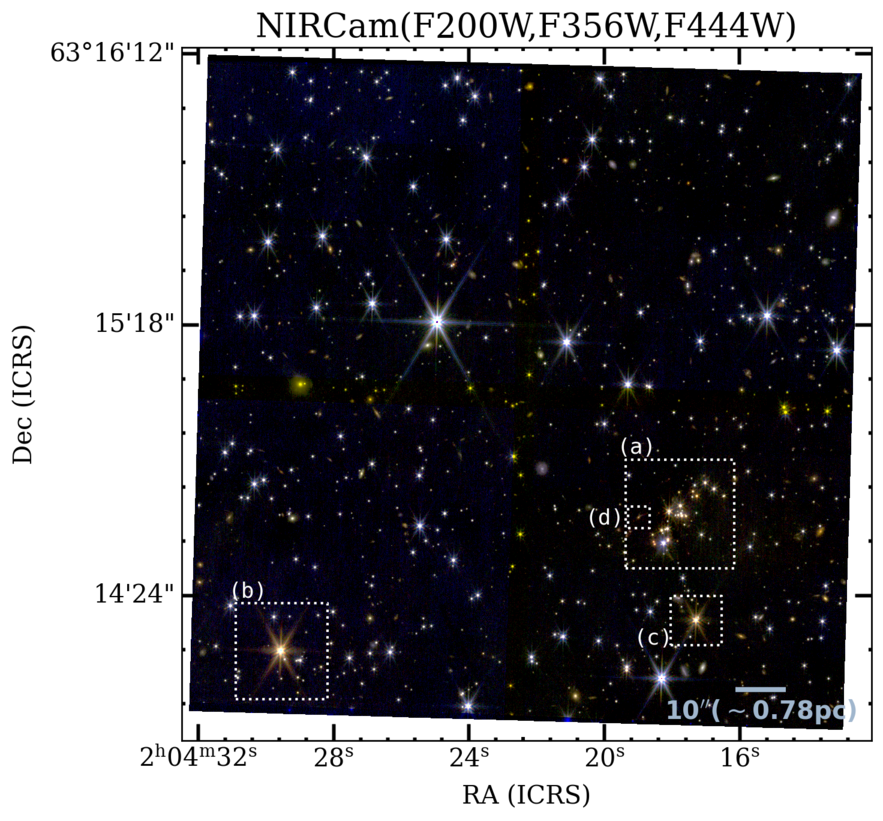}{0.65\textwidth}{}} \vspace{-9mm}
\gridline{\fig{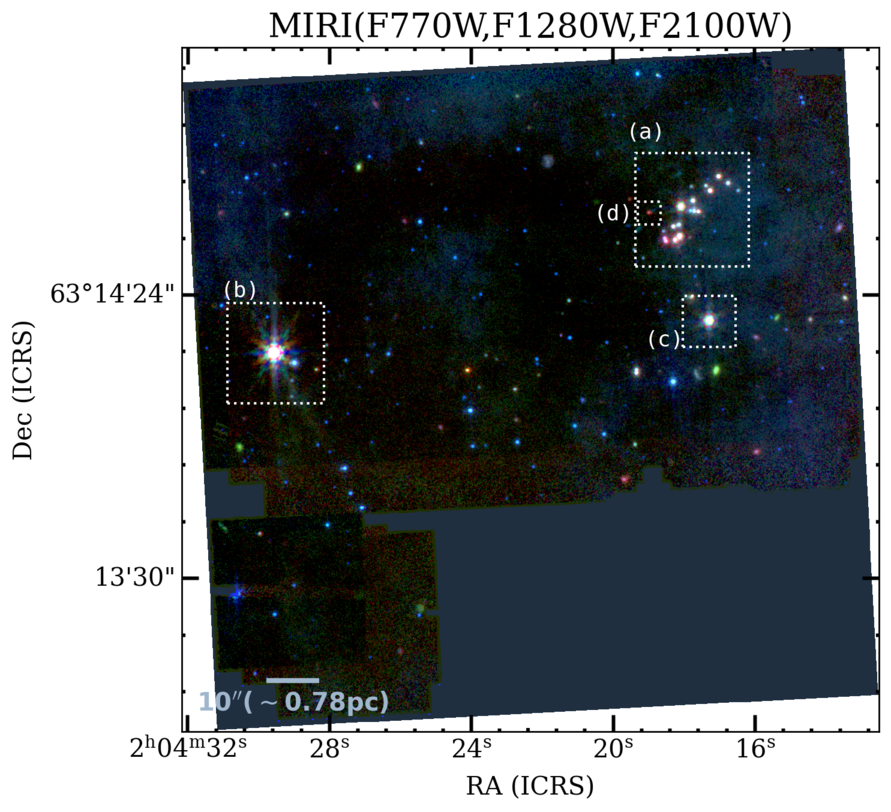}{0.65\textwidth}{} \vspace{-9mm}} 
\caption{
Top: Overall IR (2.00, 3.56, and 4.44 $\micron$) pseudocolor image of Digel Cloud 1a from NIRCam.
Bottom: Overall IR (7.7, 12.8, and 21.0 $\micron$) pseudocolor image of Digel Cloud 1a from MIRI.
The white dotted squares indicate the notable structures: (a) main cluster and (b), (c) isolated reddened single sources.
The physical scale ($\sim$ 0.78 pc) at the scale bars in both panels
is calculated assuming that the distance to Digel Cloud 1 is 16 kpc.
}
\label{entireview_dc1a} 
\end{figure*}

\begin{figure*}
\epsscale{0.9}
\plotone{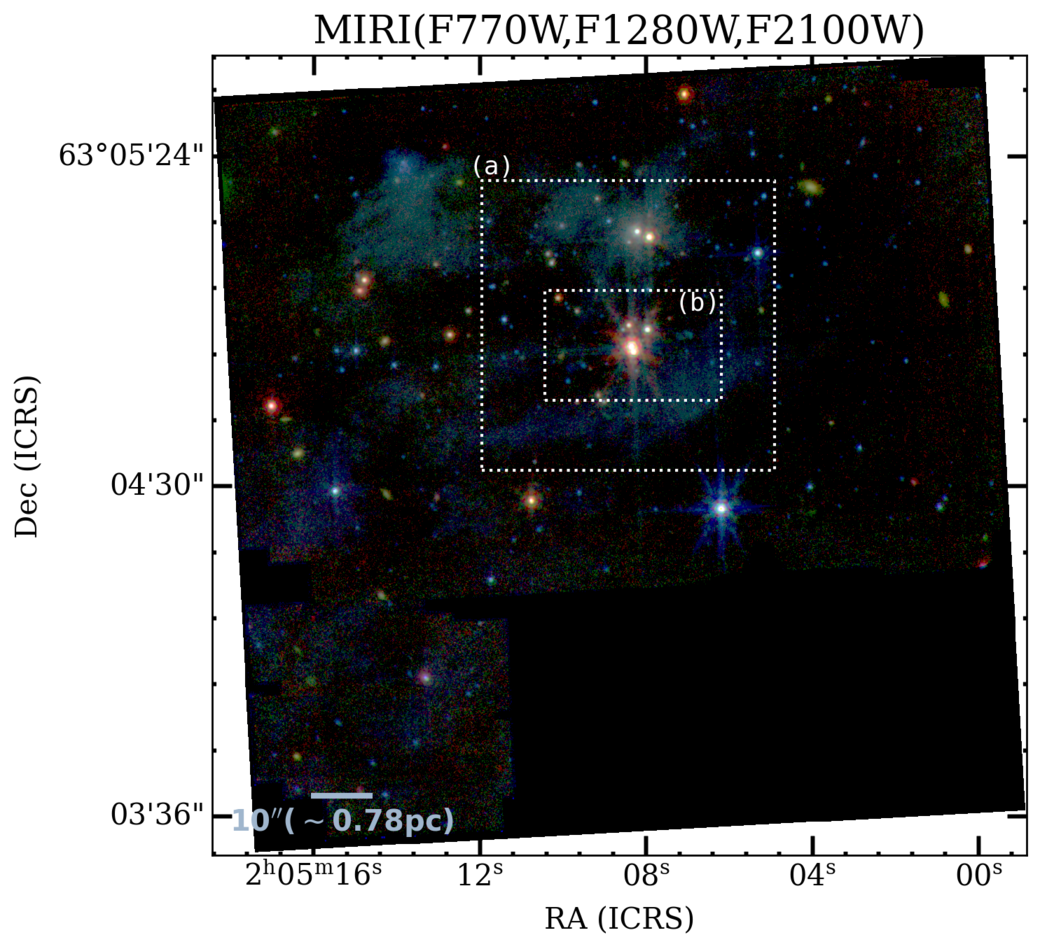}
\caption{
Overall IR (7.7, 12.8, and 21.0 $\micron$) pseudocolor image of Digel Cloud 1b from MIRI.
The white dotted squares indicate the notable structures: (a) main cluster and (b) candidates of outflow/jet components.
The physical scale ($\sim$ 0.78 pc) at the scale bars in both panels
is calculated assuming that the distance to Digel Cloud 1 is 16 kpc.
}
\label{entireview_dc1b} 
\end{figure*}

\begin{figure*}
\epsscale{0.65}
\gridline{\fig{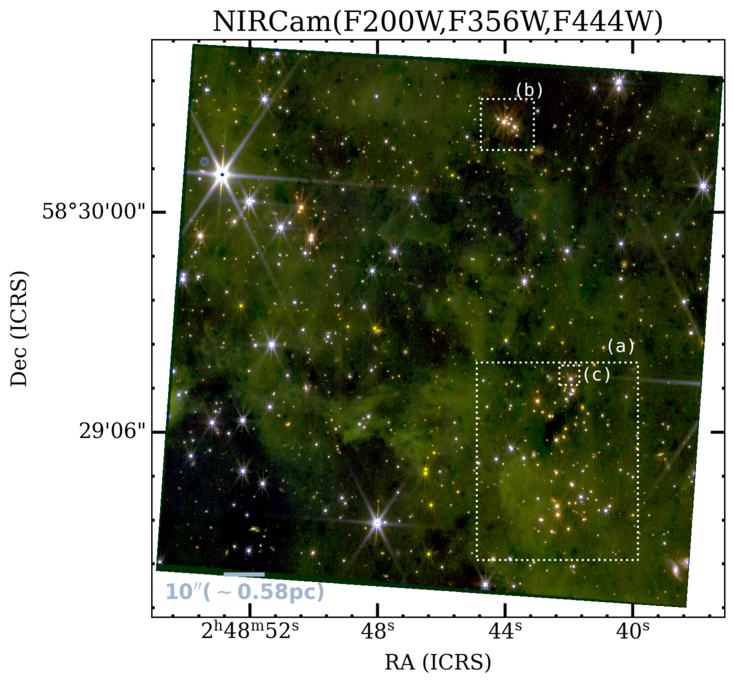}{0.6\textwidth}{}} \vspace{-9mm}
\gridline{\fig{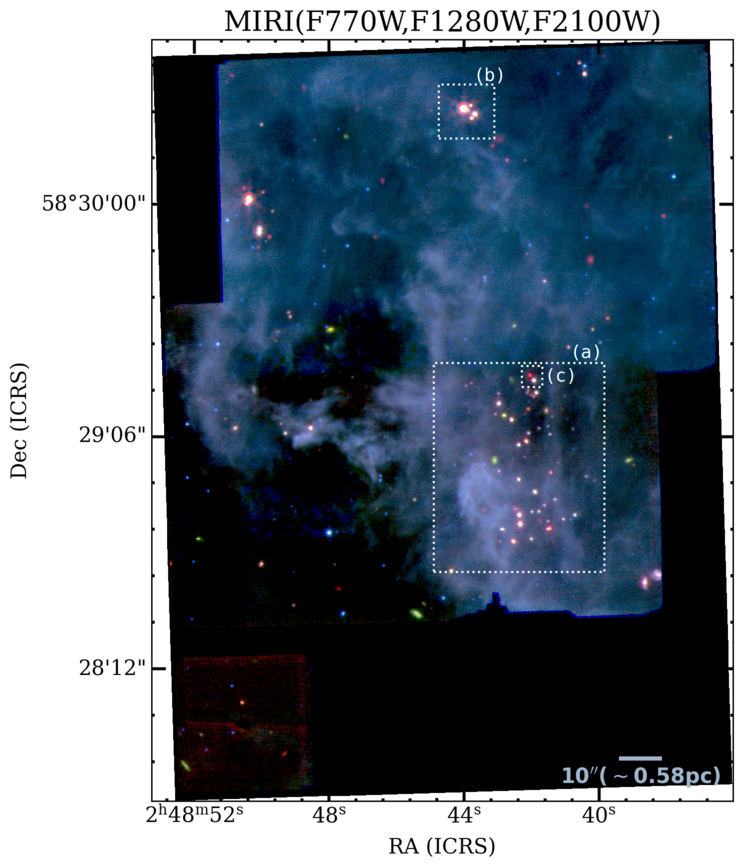}{0.6\textwidth}{} \vspace{-9mm}} 
\caption{
Top: Overall IR (2.00, 3.56, and 4.44 $\micron$) pseudocolor image of Digel Cloud 2N from NIRCam.
Bottom: Overall IR (7.7, 12.8, and 21.0 $\micron$) pseudocolor image of Digel Cloud 2N from MIRI.
The white dotted squares indicate the notable structures: (a) main cluster and (b) subcluster.
The physical scale ($\sim$ 0.58 pc) at the scale bars in both panels
is calculated assuming that the distance to Digel Cloud 2 is 12 kpc.
}
\label{entireview_dc2n} 
\end{figure*}

\begin{figure*}
\epsscale{0.65}
\gridline{\fig{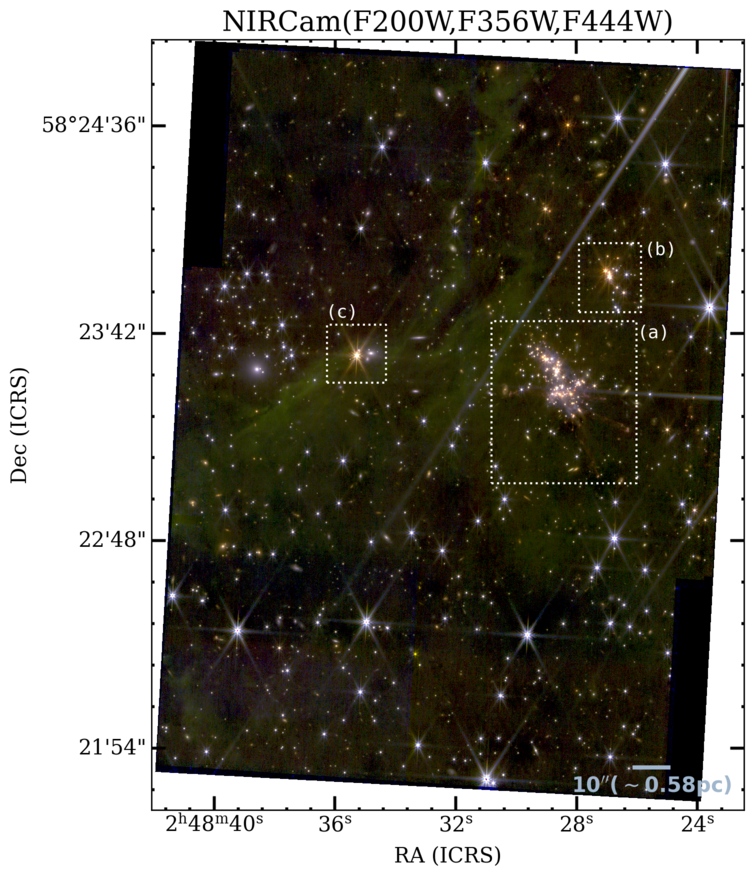}{0.6\textwidth}{}} \vspace{-9mm}
\gridline{\fig{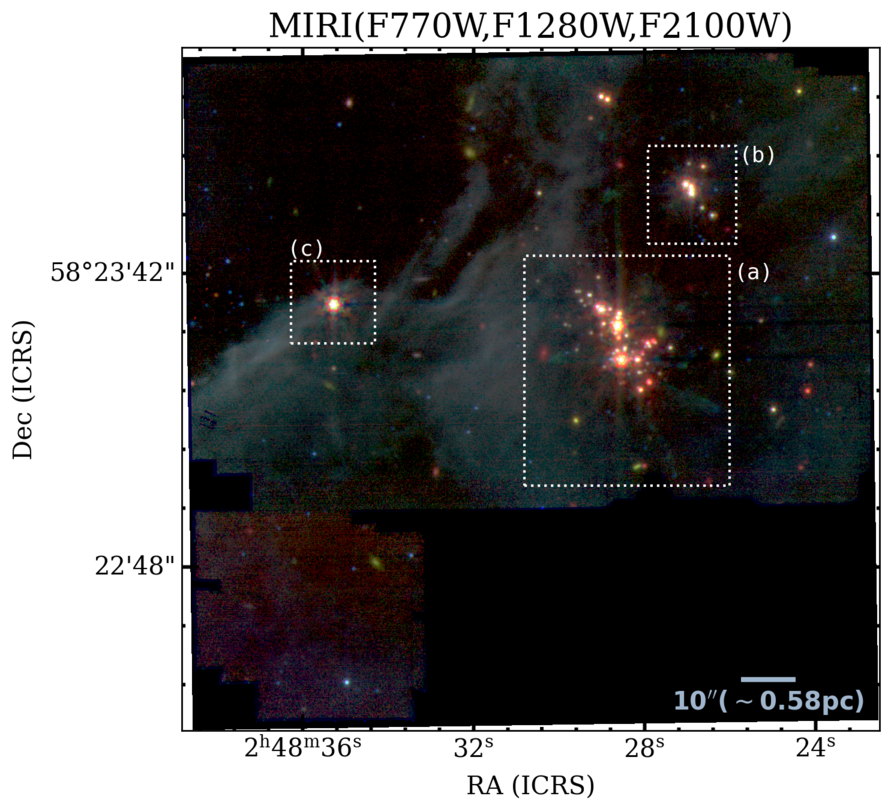}{0.6\textwidth}{} \vspace{-9mm}} 
\caption{
Top: Overall IR (2.00, 3.56, and 4.44 $\micron$) pseudocolor image of Digel Cloud 2S from NIRCam.
Bottom: Overall IR (7.7, 12.8, and 21.0 $\micron$) pseudocolor image of Digel Cloud 2S from MIRI.
The white dotted squares indicate the notable structures: (a) main cluster and (b) subcluster.
The physical scale ($\sim$ 0.58 pc) at the scale bars in both panels
is calculated assuming that the distance to Digel Cloud 2 is 12 kpc.
}
\label{entireview_dc2s} 
\end{figure*}

Figures \ref{entireview_dc1a}--\ref{entireview_dc2s} present comprehensive images of our target regions
(Figure \ref{entireview_dc1a}: Cloud 1a, Figure \ref{entireview_dc1b}: Cloud 1b,
Figure \ref{entireview_dc2n}: Cloud 2N, and Figure \ref{entireview_dc2s}: Cloud 2S)
as observed with NIRCam\footnote{Note that we did not observe Cloud 1b with NIRCam due to the aforementioned time-allocation constraints.} and MIRI.
Monochromatic images of all NIRCam and MIRI filters for all targets are shown in Appendix \ref{sec:a-2} (Figures \ref{dc1a_nircam} -- \ref{dc2s_miri}).
The main clusters are delineated by white dashed squares and are labeled as (a) in these figures. 
Within Cloud 2N and 2S, subclusters are detected near the main clusters, and are labeled as (b) in Figures \ref{entireview_dc2n} and \ref{entireview_dc2s}.
Although the presence of these subclusters was initially suggested by \citet{Izumi2017} using University of Hawaii telescope NIR data and WISE MIR data
\citep[Figures 2 and 3 in][]{Izumi2017}, the spatial resolution of the MIR data was insufficient to confirm their nature as actual clusters.
The clarity of these clusters is now evident using NIRCam and MIRI with their very high spatial resolutions.

In addition to the clusters, we have detected several structures that were not identified in previous observations.
Near the main clusters of Cloud 1a and Cloud 2S, three isolated reddened individual sources have been identified.
These sources are marked as (b) and (c) in Figure \ref{entireview_dc1a} and as (c) in Figure \ref{entireview_dc2s}.
We have also found two sources around the main clusters of Cloud 1a and Cloud 2N that exhibit much higher intensity at longer wavelengths, particularly at 21 $\micron$,
in comparison to shorter wavelengths.
These sources are likely in a very early stage of evolution, specifically the Class 0 stage \citep[e.g.,][]{Fischer2016}, and are labeled as (d) in Figure \ref{entireview_dc1a} and (c) in Figure \ref{entireview_dc2n}.
Possible outflow or jet components have been detected within the Cloud 1b and Cloud 2S clusters.
The outflows in Cloud 1b are contained within (b) in Figure \ref{entireview_dc1b}.
The outflows in Cloud 2S are located in various places around the main cluster, labeled as (a) in Figure \ref{entireview_dc2s}.
Additionally, we have confirmed several new nebular structures around the main clusters.
More detailed discussions of these notable structures are in Section \ref{sec:dis}.

There are, however, a number of instrumental artifacts in these images, too. They sometimes appear as straight features connecting to bright stars.
These are generally diffraction spikes, both from bright sources in the cluster (common in both the NIRCAM and MIRI data) and from very bright stars outside the FoV (primarily in the NIRCAM data). There is also an additional set of spikes in the MIRI data at 7.7 \micron{} due to diffraction from the gridded nature of the pixel layout in the detector \citep{Gaspar2021}. It is important to note that our
results are not affected by these artifacts because we use PSF fitting for the photometry (Section \ref{sec:data_red}), and we do not treat structures that could easily be confused with artifacts as notable structures 
in the previous paragraph.

\subsection{Mass Detection Limits}\label{sec:res-2}
We determined the mass detection limits by utilizing the measured sensitivities (Table \ref{sensitivity})
employing a model for pre-main-sequence stars available for all NIRCam and MIRI filters provided by \citet{Baraffe2015}.
This model encompasses a mass range from 0.01 $M_\odot$ to 1.4 $M_\odot$ within the age range of 0.5--10 Myr.
Beyond 10 Myr, the minimum mass in the mass range increases with age.
The adopted distances for Cloud 1 and 2 clusters are 16 kpc and 12 kpc, respectively, while the adopted ages for both Cloud 1 and 2 clusters range from 0.5 to 1.0 Myr.
These values are derived from previous studies (Table \ref{prop-tar})
\footnote{In this paper, we adopt a distance to Cloud 2 of $D$ = 12 kpc because the spectroscopic distance of stars should be more accurate than the kinematic distance \citep[see Table \ref{prop-tar}; e.g.,][]{ Kobayashi2008, Yasui2006, Yasui2008}.}.
We note that only 0.5 and 1.0 Myr models are provided in the range of 0.5 to 1.0 Myr.
Therefore, we compared these two models and selected a larger mass of the two models as a detection limit.
We also accounted for the impact of line-of-sight extinction based on our previous results
\citep[Cloud 1a: $A_V$ = 6.5 mag, 1b: $A_V$ = 5.5 mag, Cloud 2N: $A_V$ = 7.2 mag, 2S: $A_V$ = 6.1 mag;][]{Yasui2008,Izumi2014}.
The extinction law referenced in this paper is from \citet{Wang2019}, 
which predicted the average extinction law with the total-to-selective extinction ratio $R_{\rm V}$ ($=$ $A_{\rm V}/E(B - V) = A_{\rm V}/(A_{\rm B} - A_{\rm V})$) = 3.16 for various NIRCam filters with a 2.5 \% precision.
For the three MIRI filters, we did not correct for extinction because it becomes negligible at wavelengths longer than about 3 $\micron$ \citep[e.g.,][]{McClure2009,Wang2019}
\footnote{The impact around 10 $\micron$ (9.7 $\micron$ interstellar silicate extinction) is known to be comparable to or larger than those around 3 $\micron$. 
However, our selected MIRI filters (F770W, F1280W, and F2100W) are well separated from the 9.7 $\micron$ feature, allowing us to ignore its effect.}.

The estimated mass detection limits are summarized in Table~\ref{limit}.
We find that we are able to detect YSOs with masses of about 0.01--0.05 $M_\odot$ with NIRCam filters, except for F405N.
This value is about 10 times better than the previous observations \citep{Yasui2008}.
The mass detection limits of F405N, which is a narrow-band filter, reach about 0.1 $M_\odot$.
With the MIRI 770W filter, we are able to detect YSOs with masses of about 0.1--0.3 $M_\odot$.
Owing to such high sensitivity and resolution, we could detect individual YSOs in wavelengths longer than 2.0 $\micron$ in the EOG for the first time. 
Using these data, we also present a thorough investigation of the peak value of the IMF and the evolution of 
YSOs in Cloud 2 clusters in Yasui et al. (2024, submitted).

\subsection{Color-Magnitude Diagrams}\label{sec:res-3}
\begin{figure*}
\gridline{\fig{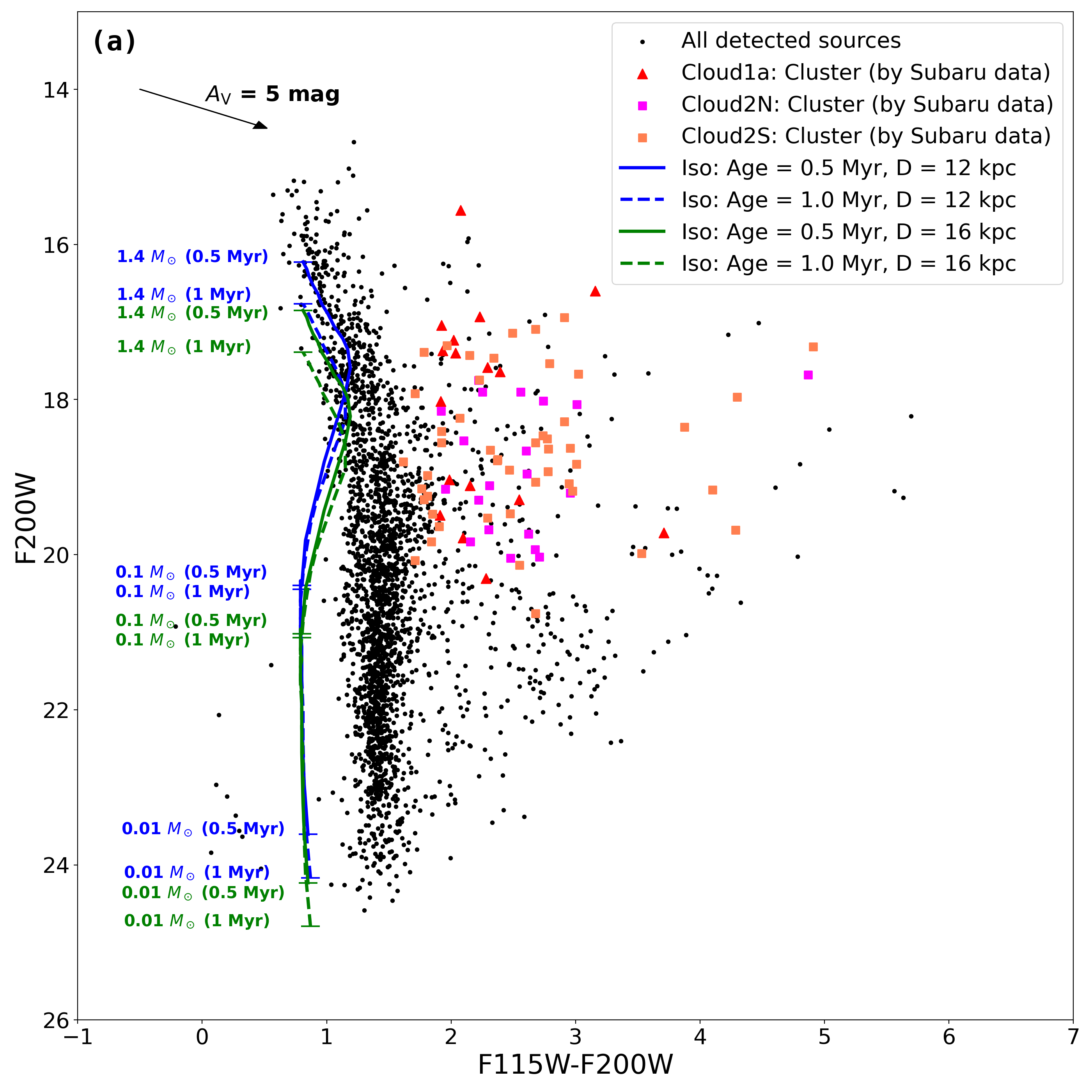}{0.37\textwidth}{}
          \fig{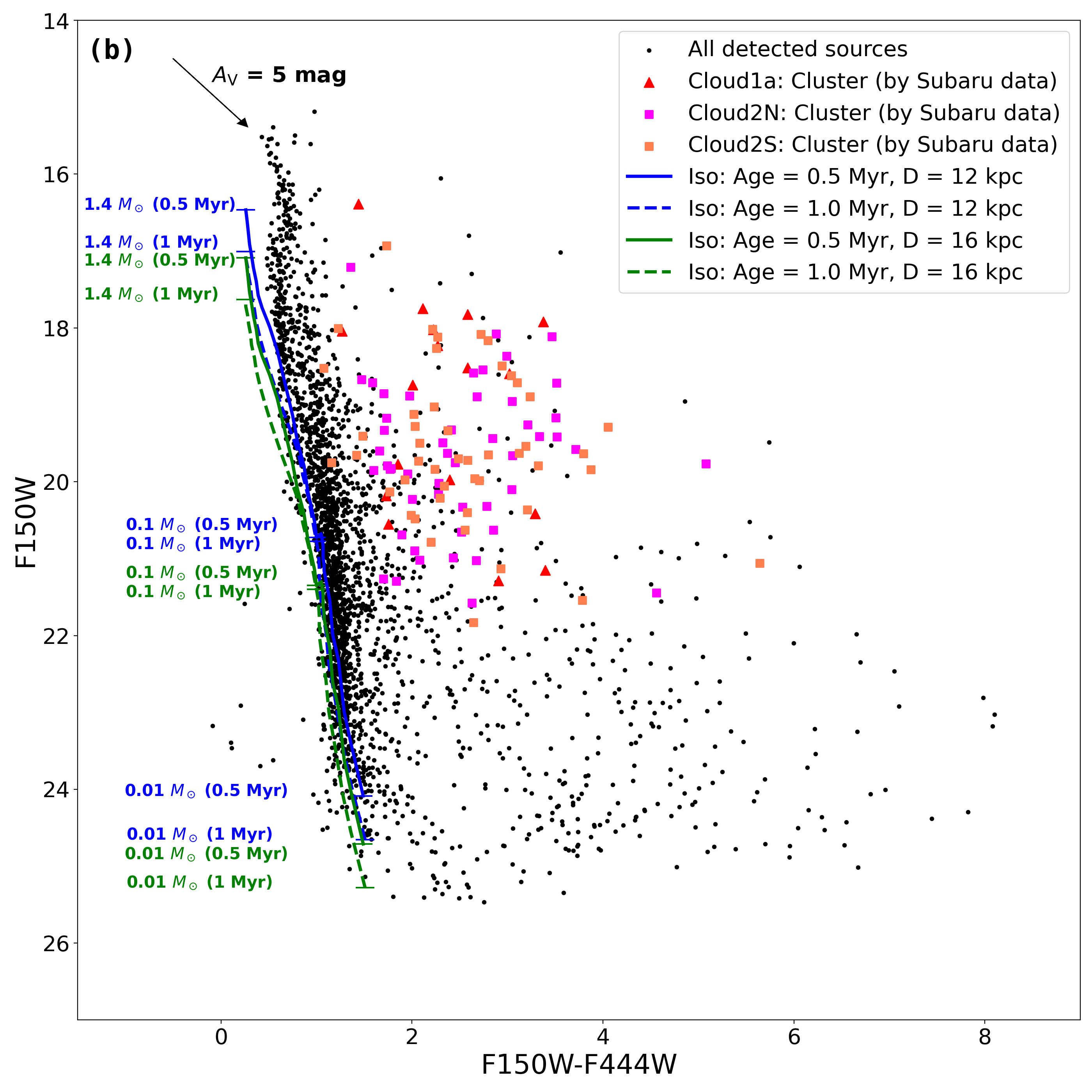}{0.37\textwidth}{}} \vspace{-9mm}
\gridline{\fig{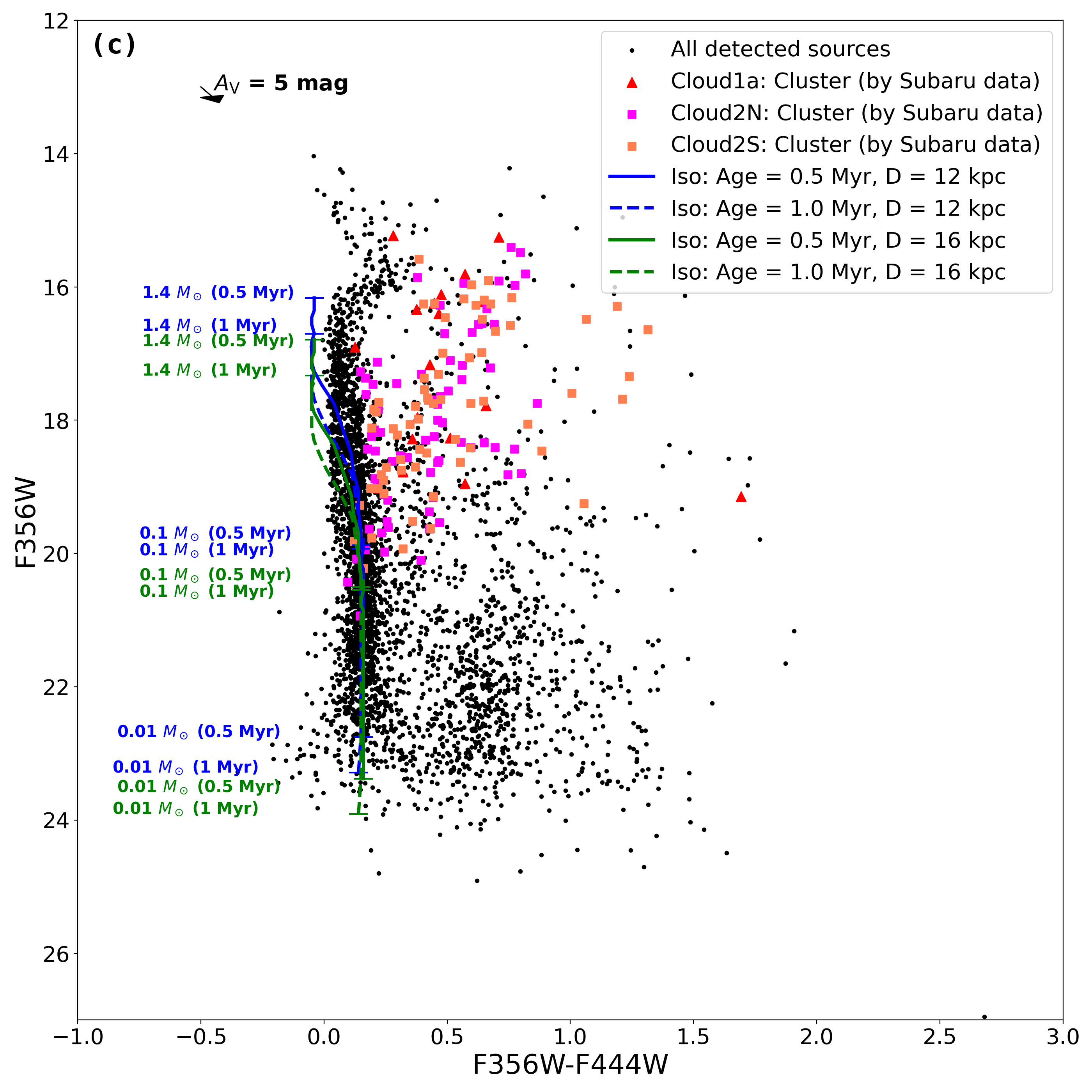}{0.37\textwidth}{}
          \fig{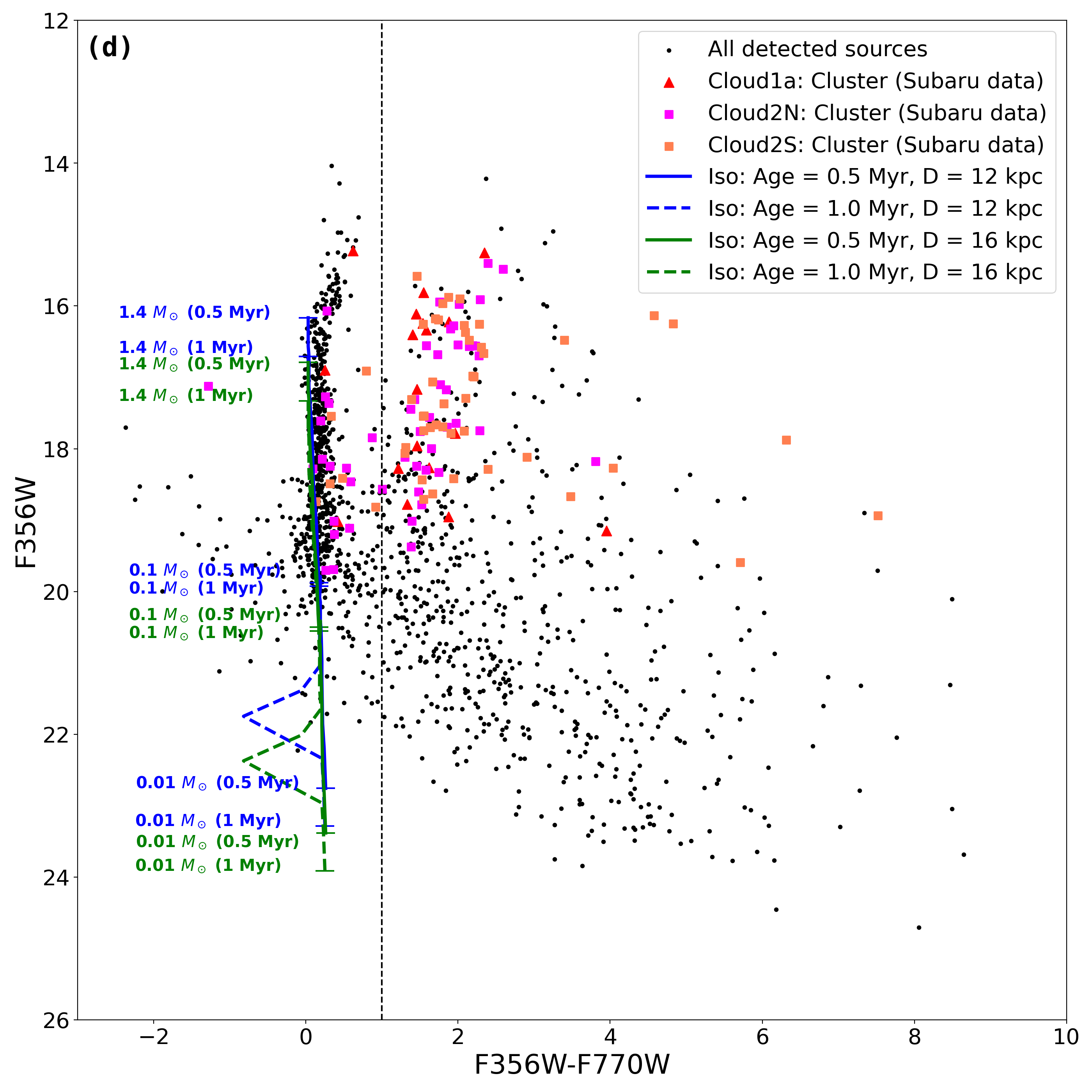}{0.37\textwidth}{}} \vspace{-9mm}
\gridline{\fig{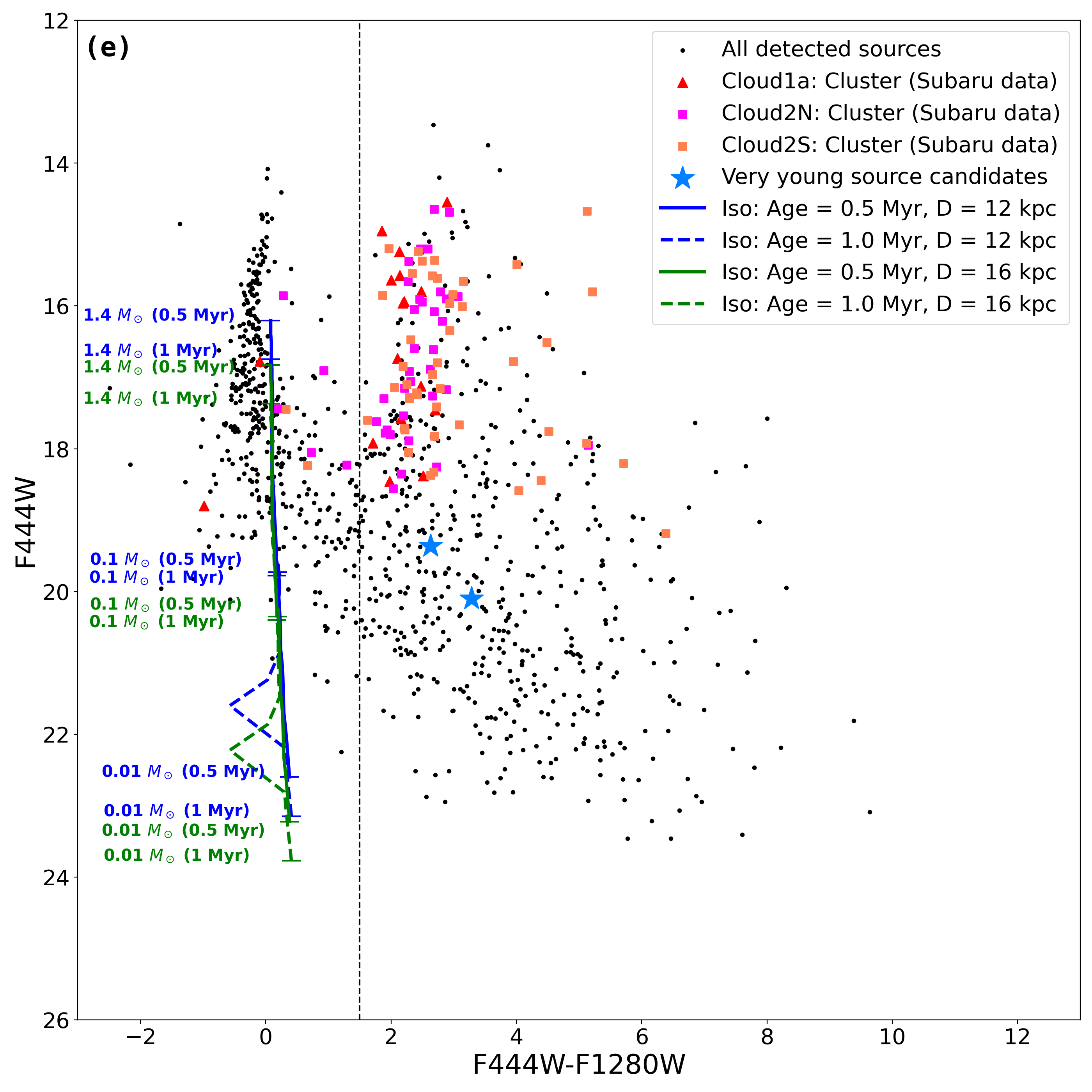}{0.37\textwidth}{}
          \fig{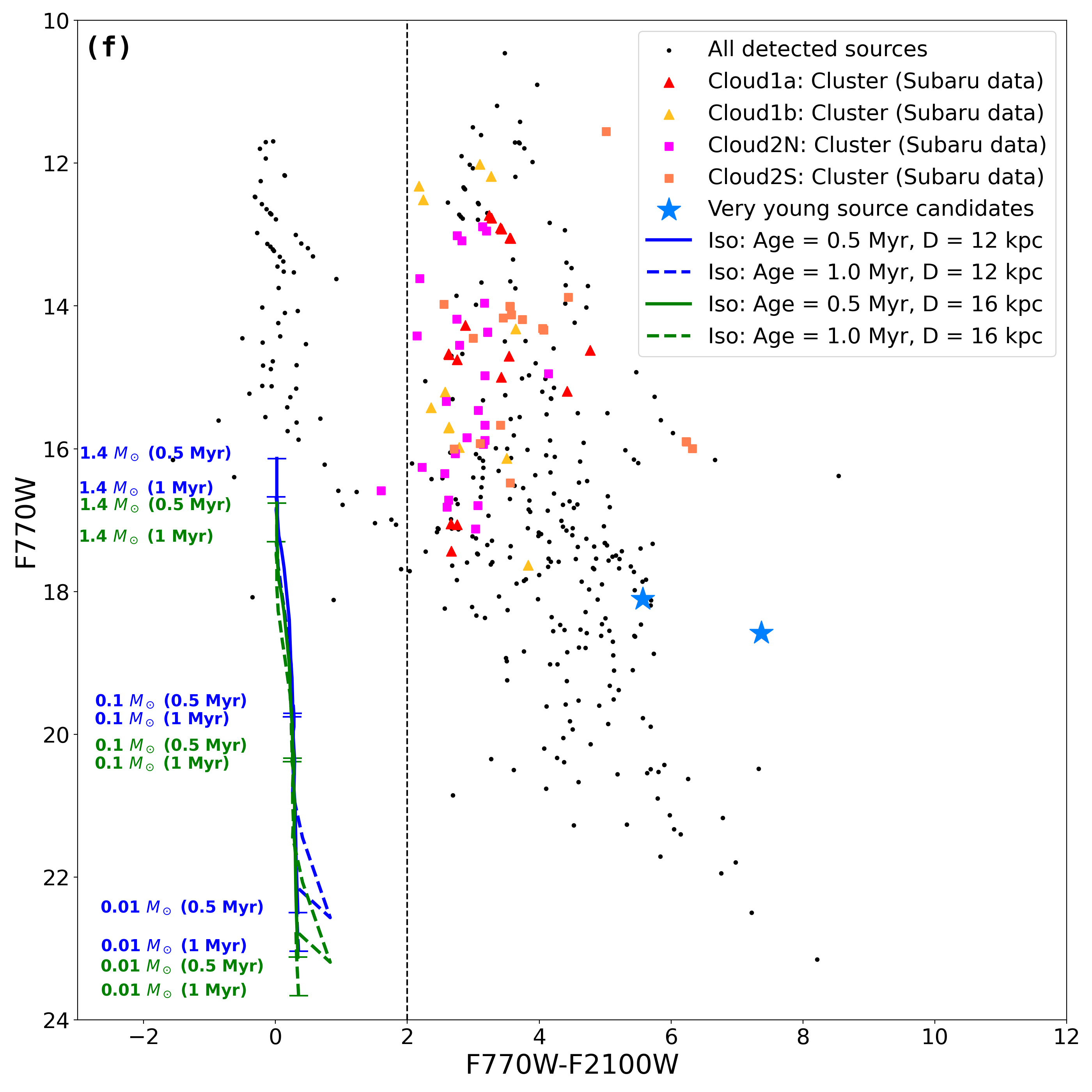}{0.37\textwidth}{}
          }
\caption{
CM diagrams for all detected sources.
Red triangles, yellow triangles, magenta squares, and orange squares indicate the cluster members within Cloud 1a, Cloud 1b, Cloud 2N, and Cloud 2S, respectively.
These cluster members were identified in Subaru data \citep{Yasui2009, Izumi2014}.
The black dots indicate all sources.
The black arrows in panels (a), (b), and (c) show the reddening vectors of $A_{\rm V}$ = 5 mag \citep{Wang2019}
The blue solid and dashed lines represent models for pre-main-sequence stars (isochrone model) by \citet{Baraffe2015} at ages of 0.5 Myr and 1.0 Myr, respectively, with a distance of 12 kpc.
Similarly, the green solid and dashed lines depict models for pre-main sequence stars (Isochrone model) by \citet{Baraffe2015} at ages of 0.5 Myr and 1.0 Myr, respectively, but with a distance of 16 kpc.
The blue stars indicate the very young source candidates.
The black dashed lines indicate the threshold between cluster members and other sources in longer-wavelength colors with the criteria: F356W-F770W = 1.0 (d), F444W-F1280W = 1.5 (e), and F444W-F2100W = 2.0 (f).
}
\label{cm} 
\end{figure*}

Figure \ref{cm} displays color-magnitude (CM) diagrams for all detected sources.
In addition to these sources, we have superimposed the pre-main-sequence models onto the diagrams as discussed in Section \ref{sec:res-2}. 
Within all CM diagrams, cluster members as previously identified in Subaru data \citep[e.g.,][]{Yasui2009,Izumi2014} occupy a redder region compared to the background sources. 
This can be attributed to intracluster extinction since the sources are still embedded in the cluster environment. 
Moreover, younger sources are known to exhibit a significant infrared excess resulting from circumstellar disks and envelopes \citep[e.g.,][]{Robitaille2006, Robitaille2007}.
Particularly at longer wavelengths (as seen in panels (d), (e), and (f) in Figure \ref{cm}), there is a clear distinction in the distribution of cluster members and other sources. 
Cluster members are prominently concentrated in regions where F356W-F770W $>$ 1.0, F444W-F1280W $>$ 1.5, and F444W-F2100W $>$ 2.0 (indicated by the black dotted lines in Figure \ref{cm}).
This underscores the effectiveness of longer wavelengths in identifying YSOs and aligns with prior studies using MIR data from Spitzer and WISE \citep[e.g.,][]{Gutermuth2008,Gutermuth2009,Koenig2012,Koenig2014,Sewilo2019}.

\section{Discussion}\label{sec:dis}
\subsection{Main Clusters}
Figure \ref{zoom_clusters} provides a comparison between NIR images obtained with JWST (right) and Subaru \citep[left;][]{Yasui2009,Izumi2014} for the main clusters, marked as (a) in Figures \ref{entireview_dc1a}--\ref{entireview_dc2s}.
Owing to the significantly increased sensitivity of JWST (see Section \ref{sec:data_red}), the number of detected sources has increased, including the identification of lower-mass stars compared with the prior observations.

In the Cloud 1a cluster, we have confirmed that YSOs exhibit an elongated distribution instead of a circular one, and they are associated with the dust features revealed by F770W data.
This elongated distribution is consistent with the expectation that YSOs would be linked to the filamentary structures \citep[][; Section \ref{sec:tar}]{Inoue2018,Tokuda2019} because the formation of the Cloud 1a cluster is thought to be triggered by the collision between Complex H and the Galactic disk \citep{Izumi2014}.

The Cloud 2N cluster has a larger size than the other clusters (Cloud 2N: radius $\simeq$ 0.6 pc, Cloud 2S: radius $\simeq$ 0.3 pc, Cloud 1a: radius $\simeq$ 0.4 pc).
Its morphology is that of a loose association.  Cloud 2S, on the other hand, is a concentrated, dense cluster.
This result is consistent with the previous ground-based studies \citep[][see Section \ref{sec:tar};]{Kobayashi2008,Yasui2008}.
Furthermore, the southern part of the Cloud 2N cluster is embedded in dust, as indicated by the F770W, F1280W, and F2100W images, 
while the northern part of the cluster is not obscured by dust (bottom panel of Figure \ref{entireview_dc2n}).

The F115W, F150W, and F200W images of the Cloud 2S cluster (bottom right panel of Figure \ref{zoom_clusters}) reveal extended nebulous features in addition to the stars.
Similar structures are not detected in the other clusters that were imaged with the NIRCam filters,
nor do the Subaru data reveal similar features due to limitations in resolution and sensitivity. 
Such short-wavelength extended structures have often been found in high-mass star-forming regions such as the nebular clusters like the Orion Nebular Cluster, RCW 38, and NGC 7758 \citep[e.g.,][]{Ojha2004,Wolk2006,Muench2008}.
However, we note that high-mass stars (early B- or O-type stars) have not been detected in the Cloud 2S cluster.

\subsection{Isolated Reddened Single Sources}
Within Cloud 1a, the separation between the isolated reddened single sources and the main cluster is approximately 85\arcsec{} (labeled as (b) in Figure \ref{entireview_dc1a}),
equivalent to 7.4 pc, and 20\arcsec{} (labeled as (c) in Figure \ref{entireview_dc1a}), equivalent to 1.6 pc. 
Both stars are located at the peak of the Cloud 1a molecular cloud, as identified through $^{12}$CO(1-0) observations conducted with the NRO 45m telescope \citep{Izumi2014},
and exhibit a similar structure and brightness as the main cluster in the WISE data. 
These sources were previously interpreted as small aggregations composed of intermediate-mass stars associated with 
Cloud 1a in the previous studies \cite[e.g.,][]{Izumi2014,Izumi2017};
however, with the high spatial resolution and sensitivity of JWST images, it becomes evident that these are not aggregations but rather isolated single stars.
This finding suggests that both isolated higher-mass star formation and complex star formation in cluster mode can occur within the same molecular cloud.

The isolated source near the Cloud 2S main cluster, known as IRS4, was also  thought to form a typical small aggregation around intermediate-mass stars
\citep{Kobayashi2008}.
It was identified as a candidate star-forming region in \citet{Izumi2017}, as shown in panel (xv) in their Figure 3.
In that figure, several red objects appear in the NIR image from Subaru, while unresolved objects seem to be present in the WISE data
\citep[labeled as J024835.25+582336.1 in the AllWISE catalog;][]{https://doi.org/10.26131/irsa1}.
However, the high-spatial-resolution JWST images reveal that it is a single isolated star, surrounded by background galaxies.

\subsection{Potentially Very Young Sources}
Figure \ref{can_class0} displays NIRCam and MIRI images of two potentially very young sources (top: labeled as (d) in Figure \ref{entireview_dc1a}, bottom: labeled as (c) in Figure \ref{entireview_dc2n}). 
Despite the better sensitivities of the F115W and F150W filters compared to the longer-wavelength filters (Section \ref{sec:data_red}, Table \ref{sensitivity}), these sources are undetected at the shortest wavelengths.
These very reddened sources may be in a very early stage of evolution, possibly the class 0 stage \citep[e.g.,][]{Fischer2016}.

\citet{Fischer2016} established WISE color criteria for Class 0 protostars using color-color diagrams W1-W2 versus W2-W3 and W2-W3 versus W3-W4. 
The criteria include: (1) W1-W2 $>$ 1 and 
(2) W2-W3 $<$ 1.8 $\times$ (W3-W4) - 6.5
(see Figures 11 and 12, Equations (11) and (12) in \citet{Fischer2016}).  
To assess these criteria, we examined the color-color diagrams with the nearest equivalent wavelength filters on JWST:
F356W-F444W versus F444W-F1280W (left panel of Figure \ref{cc})
and F444W-F1280W versus F1280W-F2100W (right panel of Figure \ref{cc}).
The criteria are overlaid on the color-color diagrams (black dashed lines in Figure \ref{cc}). 
We confirmed that one of the sources, which is located in Cloud 2N cluster, meets both criteria (1) and (2). 
The other source, which is located in Cloud 1a cluster, meets criterion (1) but narrowly misses criterion (2).
However, this source (located in Cloud 1a) is closer to the criterion than the other cluster members.
Therefore, both sources are the most probable class 0 candidates among those detected.

As these objects were not identified in previous observations, this represents their first detection enabled by the high sensitivity, long-wavelength coverage of MIRI. 
By applying the YSO identification and classification criteria outlined in \citet{Koenig2014},
we can rule out the possibility that these sources are AGB stars or star-forming galaxies (e.g., as depicted in Figures 5 in \citet{Koenig2014}, and Figure 2 in \citet{Fischer2016}).
Nevertheless, we cannot dismiss the possibility that these objects may be foreground/background planetary nebulae, background active galactic nuclei, 
or quasistellar objects, as these types of objects also exhibit similar colors \citep[e.g.,][]{Wright2010,Koenig2014}.

We note that a few cluster members are also closer to class~0 criterion (2), as seen in the right panel of Figure \ref{cc}.
However, in these cases, these members are positioned very close to nearby bright sources.
Consequently, due to the larger spatial resolution at F2100W, these sources, which are clearly detected in shorter-wavelength filters, cannot be detected as individual sources
(i.e., they are detected as one source combined with nearby bright sources) in F2100W.
Such sources exhibit artificially elevated F2100W photometry. Therefore, visual inspection of all sources is crucial to exclude those with inflated F2100W or other filter photometries.

\subsection{Possible Outflow or Jet Components}
We have identified several bow-shock and knot structures within the main clusters of Cloud 1b and Cloud 2S based on their resemblance to confirmed jets or outflows in nearby star-forming regions \citep[e.g.,][]{Velusamy2011, Ray2023}.
Figure \ref{zoom_outflows} presents a comparison between pseudocolor images obtained with JWST (right) and previous data from WISE (top left) or Spitzer (bottom left) for these structures (top: Cloud 1b, bottom: Cloud 2S). 

In the case of the Cloud 1b cluster, pairs of bow-shock-like structures detected by the 770W and 1280W filters are situated around the bright central star in the cluster
(top-right panel of Figure  \ref{zoom_outflows}). 
These structures are distinguishable from background galaxies based on their color, and each structure appears on the opposite side of the star.
As such, they are likely components of a jet head.
Candidate knot structures are also detected near the jet head located on the western side of the star (highlighted by a white dashed square in the top right panel of Figure \ref{zoom_outflows}). 
Based on the previous studies of outflow/jet structures with MIRI \citep[e.g.,][]{yang2022},
it is possible that these structures are dominated by H$_2$~0-0~S(5) and H$_2$~0-0~S(4) in the F770W filter, and H$_2$~0-0~S(2) and [Ne~II]~$^2P^0_{1/2} - ^2P^0_{3/2}$ in the F1280W filter.
While [Ne II] emission is more often found in collimated jet-like 
structures, H$_2$ emission generally traces a more extensive, wide-angle outflow cavity \citep[e.g.,][]{Bally2016,yang2022}. Spectroscopy will be needed to confirm whether these are indeed H$_2$ flows.

Figure \ref{dc1b_detail_outflow} illustrates the comparison between F770W and F1280W images for the potential jet head located on the western side of the star. 
The F770W filter reveals a double-peaked structure, while the F1280W filter shows an arc-like configuration. 
The knot structure also stands out much more clearly in the F770W filter.
The projected distance from this presumed jet head to the central star is approximately 1.5 $\times$ 10$^{5}$ AU (from the western side to the star)
and 1.8 $\times$ 10$^{5}$ AU (from the eastern side to the star). 
The projected distance from the potential knot structure to the central star is approximately 1.2 $\times$ 10$^{5}$ AU.

Several bow-shock and knot structures are evident at various locations within the Cloud 2S main cluster. 
These structures are discernible not only in the MIRI filters (F770W and F1280W) but also in the NIRCam filters (F200W, F356W, and F444W). 
Figure \ref{dc2s_detail_outflow} provides a comparative view of the F200W, F356W, F444W, F770W, and F1280W images, showcasing two representative outflow or jet candidates (highlighted by a white dashed square in the bottom right panel of Figure \ref{zoom_outflows}).
Based on the previous studies of outflow/jet structures with NIRCam \citep[e.g.,][]{Ray2023},
we speculate that these structures detected by NIRCam may be linked to emission from H$_2$ 1-0 S(1) in F200W,
H$_2$ 1-0 O(5) in F356W, H$_2$ 0-0 S(9), and CO (1-0) rotational-vibrational emission in the F444W filter.
Figure \ref{dc2s_detail_outflow} provides a comparative view of the F200W, F356W, F444W, F770W, and F1280W images,
showcasing two representative potential outflow or jet components (highlighted by a white dashed square in the bottom right panel of Figure \ref{zoom_outflows}).

In the northern component (labeled as ``Candidate 1" in Figure \ref{dc2s_detail_outflow}), we observe knots and jet head structures in all filter images. 
The jet head structure in the F1280W filter image appears larger and brighter than the knot structure though other filters show no significant distinction.
In contrast, in the southern component, the jet head structure, as traced by the F1280W filter, is smaller and weaker than the knot component traced by the same filter and the jet head structure traced by other filters. 
Furthermore, in the
southwestern part (labeled as ``Candidate 2" in Figure \ref{dc2s_detail_outflow}), two knot components are detected in F356, F444W, and F770W filters, while only one knot (the one closest to the jet-head structure) is detected in F200W and F1280W filters.
Moreover, the Cloud 2S main cluster has the highest source density among all our observed targets, 
further complicating the task of determining the central star or stars responsible for ejecting the various outflows.

The lower-limit projected distance from the jet head structure in the northern (Candidate 1) and southwestern (Candidate 2) portions
is estimated to be approximately 9.5 $\times$ 10$^{4}$ AU and 1.7 $\times$ 10$^{5}$ AU, respectively. 
The projected distance between the knot and the jet-head structure in the northern region is roughly 1.2 $\times$ 10$^{4}$ AU.
The projected distance between the jet-head structure, the knot nearest to it, and the other knot in the southwestern region measures approximately 1.6 $\times$ 10$^{4}$ AU and  6.7 $\times$ 10$^{4}$ AU.

\subsection{Nebular Features}
The nebular features, primarily detected at longer wavelengths, particularly with the F770W and F1280W filters, are observed in and around all the main clusters. 
Notably, distinct nebular structures are identified within Cloud 2N and 2S.
Figure \ref{nebula} illustrates a comparison between pseudocolor images obtained with JWST (right: F356W, F444W, and F770W filters) and previous data from Spitzer (left: IRAC ch1, ch2, and ch3 bands) for Cloud 2N (top) and 2S (bottom), providing enhanced clarity of the nebular structures. 
Because of JWST's greater spatial resolution, detailed features in the nebular structures in Cloud 2N and 2S are highlighted, allowing us to better see the complex nature of the molecular clouds.

The nebulosity within Cloud 2N displays multiple pillar- or cliff-like structures, reminiscent of those found in closer star-forming regions \citep[e.g.,][]{Koenig2012,Reiter2022}. 
In addition to the F770W and F1280W filters, the F356W filter, known to trace polycyclic aromatic hydrocarbons (PAHs), also reveals a similar structure
(see the top panel of Figure \ref{entireview_dc2n}).
It is likely that these structures form due to the influence of ultraviolet radiation emitted by the nearby B-type star, MR 1, that lies within the vicinity of the Cloud 2N main cluster (see Section \ref{sec:tar} and the bottom right panel of Figure \ref{FoV}).
This aligns with the concept that star formation in Cloud 2N was
triggered by the nearby huge SNR and by MR~1,
as described in Section \ref{sec:tar} and Figure 9 of \citet{Kobayashi2008}.

Conversely, the nebular structure within Cloud 2S displays a filamentary distribution.
Cloud 2N's main clusters and subclusters appear not associated with this filament.
The filament extends from the southeast to the northwest, covering a span of approximately 3--4 pc. 
This distinction in structure between Cloud 2N and 2S might be attributed to their distance from the \ion{H}{2} region.

Figure \ref{nebula} also shows the $^{12}$CO (1-0) gas distribution detected by the NRO 45m telescope (cyan contours).
The nebulae in both Cloud 2N and 2S align well with the more closely spaced $^{12}$CO contours (eastern part of the cloud).
These distributions suggest either increased densities or higher temperatures at the eastern part of the cloud. 
This, in turn, implies the occurrence of shocks, 
as discussed in the scenario in \citet{Kobayashi2008}, where star formation is induced by compression from the expanding SNR shell (see Section \ref{sec:tar}).

\section{Summary}\label{sec:sum}
In this paper, we have presented an overview of the first results of NIRCam and MIRI imaging observations toward two clusters located in the EOG. Our main results are as follows.
\begin{enumerate}
    \item The 10$\sigma$ limiting magnitudes of the NIRCam short filters (F115W, F150W, and F200W) are approximately 24.0--25.0 mag. 
    These values are about 10--80 times better than our previous NIR observations with Subaru.
    Based on the sensitivity and models of pre-main-sequence stars provided by \citet{Baraffe2015}, 
   we estimate the mass detection limits in the NIRCam filters to be about 0.01-0.05 $M_\odot$, which is about 10 times better than the previous observations.
   The mass detection limit for MIRI F770W reaches about 0.1--0.3 $M_\odot$. 
   Owing to this high sensitivity and resolution, we could detect individual YSOs in wavelengths longer than 2.0 $\micron$ in the EOG for the first time.

    \item CM diagrams for all detected sources show that longer wavelength filters, including F356W, F444W, F770W, F1280W,
    and F2100W, are very useful for distinguishing cluster members and other sources. In particular, we confirmed that the cluster members are concentrated in the regions
    where F356W-F770W $>$ 1.0, F444W-F2100W $>$ 1.5, and F444W-F2100W $>$ 2.0.

    \item Owing to the much higher sensitivity and spatial resolution, we could detect several structures that were not identified in previous observations:
    class 0 candidates, possible outflows or jets, and distinctive nebular structures around and within the clusters.
\end{enumerate}

In future work, following up on this overview paper, we will further identify cluster members based on these JWST data, investigate their spatial distributions,
and begin to incorporate multiple-wavelength data, including submillimeter wavelengths.

\begin{acknowledgments}
This research is based on observations made with the NASA/ESA Hubble Space Telescope obtained from the Space Telescope Science Institute, which is operated by the Association of Universities for Research in Astronomy, Inc., under NASA contract NAS 5-26555. 
The observations are associated with JWST GTO Cycle 1 program ID 1237.
The JWST data used in this paper can be found in MAST:\dataset[10.17909/a2k0-tf62]{http://dx.doi.org/10.17909/a2k0-tf62}.
NI and PMK acknowledge support from the National Science and Technology Council (NSTC) in Taiwan through grants
NSTC 112-2112-M-001-049 -, NSTC 111- 2112-M-001-070-, NSTC 110-2112-M-001-057-, and NSTC 113-2112-M-001-016.
C.Y. is supported by KAKENHI (18H05441) Grant-in-Aid for Scientific Research on Innovative Areas.
NI and RL thank Christina Williams for the valuable discussions on NIRCAM photometry.
NI and CY thank Takahiro Morishita for the valuable discussions on data reduction.
NI thanks Morten Andersen for valuable discussions on photometry.
The work of MER was carried out at the Jet Propulsion Laboratory, California
Institute of Technology, under a contract with the National Aeronautics and
Space Administration.
This work is based on observations made with the NASA/ESA/CSA James Webb Space Telescope. The data were obtained from the Mikulski Archive for Space Telescopes at the Space Telescope Science Institute, which is operated by the Association of Universities for Research in Astronomy, Inc., under NASA contract NAS 5-03127 for JWST. These observations are associated with program \#1237.
This work is based [in part] on observations made with the Spitzer Space Telescope, which is operated by the Jet Propulsion Laboratory, California Institute of Technology under a contract with NASA.
This publication makes use of data products from the Wide-field Infrared Survey Explorer, which is a joint project of the University of California, Los Angeles, and the Jet Propulsion Laboratory/ California Institute of Technology, funded by the National Aeronautics and Space Administration.
This research has made use of the NASA/IPAC Infrared Science Archive, which is funded by the National Aeronautics and Space Administration and operated by the California Institute of Technology. 
This research made use of Montage.
It is funded by the National Science Foundation under grant No. ACI-1440620, and was previously funded by the National Aeronautics and Space Administration’s Earth Science Technology Office, Computation Technologies Project, under Cooperative Agreement Number NCC5-626 between NASA and the California Institute of Technology.
The research presented in this paper has used data from the Canadian Galactic Plane Survey, a Canadian project with international partners, supported by the Natural Sciences and
\end{acknowledgments}
\begin{acknowledgments}
Engineering Research Council.
This research used the facilities of the Canadian Astronomy Data Centre operated by the National Research Council of Canada with the support of the Canadian Space Agency.
This research has made use of "Aladin sky atlas" developed at CDS, Strasbourg Observatory, France.
\end{acknowledgments}

\begin{deluxetable*}{cccccccc}
\tablecaption{Target Properties.  \label{prop-tar}}
\tablewidth{0pt}
\tablehead{
\colhead{Target} & \colhead{Coordinate} & \colhead{Cloud Mass} & \colhead{Number \tablenotemark{a}} & \colhead{Age} 
& \colhead{Kinematic $R_{\rm G}$($D$)} & \colhead{Photometric $R_{\rm G}$ ($D$)\tablenotemark{b}} & \colhead{References}\\
                 & \colhead{(Ra, Dec)}  & \colhead{(10$^3$ $M_\odot$)} &   \colhead{of stars}      & \colhead{(Myr)} 
& \colhead{(kpc)}                   &\colhead{(kpc)}                    &\\
}
\startdata
Cloud 1a & 02h04m17.8s, +63d14m39s & 3.0  &  18 & $<$ 1 & 22 (16) & $\geq$ 19 ( $\geq$ 12) & (1),(2)\\ 
Cloud 1b & 02h05m8.6s, +63d04m54s   & 3.5  &  45 & $<$ 1 & 22 (16) & $\geq$ 19 ( $\geq$ 12)& (1),(2)\\  \hline
Cloud 2N & 02h48m42.0s, +58d29m3s  & 8.5  &  66 & 0.5-1.0 & 23.6 (17) & 19 (12)    & (2),(3),(4),(5),(6),(7)\\ 
Cloud 2S & 02h48m28.6s, +58d23m30s  & 14   &  72 & 0.5-1.0 & 23.6 (17) & 19 (12)  & (2),(3),(4),(5),(6),(7)\\ 
\enddata
Notes.
\tablenotetext{a}{The number of stars is derived from Subaru 8.2 m observations.}
\tablenotetext{b}{The photometric distance of Cloud 1 is estimated from the K-band luminosity function (see details in \citet{Izumi2014},
while the photometric distance of Cloud 2 is estimated from high-resolution optical spectroscopy of the B-type star MR 1 \citep{Smartt1996}
based on non-LTE model stellar atmospheres.
MR 1 is apparently associated with Cloud 2 \citep{deGeus1993}.}
References:
(1) \citet{Izumi2014}; (2) \citet{Digel1994}; (3) \citet{Stil2001}; (4) \citet{Kobayashi2008};
(5) \citet{Yasui2006}; (6) \citet{Yasui2008}; 
(7) \citet{Yasui2010}.
\end{deluxetable*}
\begin{deluxetable*}{cccccccccc}
\tablecaption{Sensitivities.\label{sensitivity}}
\tablewidth{0pt}
\tablehead{
\colhead{Target} 
& \colhead{F115W} & \colhead{F150W} & \colhead{F200W} 
& \colhead{F356W} & \colhead{F444W}  & \colhead{F405N} 
& \colhead{F770W} & \colhead{F1280W} & \colhead{F2100W}\\
& \colhead{(mag)} & \colhead{(mag)} &  \colhead{(mag)} 
& \colhead{(mag)} & \colhead{(mag)}  &\colhead{(mag)} 
& \colhead{(mag)} & \colhead{(mag)} & \colhead{(mag)}
}
\startdata
Cloud 1a 
& 24.9  & 24.7  & 24.3 
& 23.5  & 22.8  & 19.8 
& 19.0  & 17.1  & 14.1  \\
Cloud 1b
& --- & --- & --- 
& --- & --- & --- 
&19.0 & 17.1  & 14.1  \\
Cloud 2N
& 25.0  & 24.8  & 24.4 
& 23.6  & 22.9  & 19.9 
& 19.2  & 17.3  & 14.3 \\
Cloud 2S 
& 25.1  & 25.0  & 24.6 
& 23.7  & 23.0  & 20.0 
& 19.1  & 17.1  & 14.2 \\
\enddata
\end{deluxetable*}
\begin{deluxetable*}{ccccccccc}
\tablecaption{FWHM of the PSF.\label{fwhm}}
\tablewidth{0pt}
\tablehead{
\colhead{F115W} & \colhead{F150W} & \colhead{F200W} 
& \colhead{F356W} & \colhead{F444W}  & \colhead{F405N} 
& \colhead{F770W} & \colhead{F1280W} & \colhead{F2100W}\\
\colhead{(arcsec)}  & \colhead{(arcsec)} &  \colhead{(arcsec)} 
& \colhead{(arcsec)}  & \colhead{(arcsec)} &  \colhead{(arcsec)} 
& \colhead{(arcsec)}  & \colhead{(arcsec)} &  \colhead{(arcsec)} 
}
\startdata
0.066 & 0.070  & 0.080
& 0.155 & 0.173  & 0.165
& 0.312 & 0.459  &  0.730 \\
\enddata
\end{deluxetable*}
\begin{deluxetable*}{cccccccccccc}
\tablecaption{Mass Detection Limits.\label{limit}}
\tablewidth{0pt}
\tablehead{
\colhead{Target} & \colhead{F115W} & \colhead{F150W} & \colhead{F200W} & \colhead{F356W} & \colhead{F444W} & \colhead{F405N} & \colhead{F770W} & \colhead{F1280W} & \colhead{F2100W}\\
                 & \colhead{($M_\odot$)}  & \colhead{($M_\odot$)} &  \colhead{($M_\odot$)} & \colhead{($M_\odot$)}  & \colhead{($M_\odot$)} &  \colhead{($M_\odot$)}
                 & \colhead{($M_\odot$)}  & \colhead{($M_\odot$)} &  \colhead{($M_\odot$)}}
\startdata
Cloud 1a 
& 0.05 & 0.03 & 0.02 & 0.02 & 0.02 & 0.14 & 0.3 & $>$ 1.4\tablenotemark{a} & $>$ 1.4\\
Cloud 1b 
& ---   & ---   &  ---  & ---   & ---   & --- &  0.3 & $>$ 1.4 & $>$ 1.4 \\
Cloud 2N
& 0.04  & 0.02 & 0.01  & $<$ 0.01  & 0.01 & 0.09 & 0.1 & 0.8 & $>$ 1.4\\
Cloud 2S
& 0.03 & 0.01  & 0.01  & $<$ 0.01  & 0.01 & 0.09 & 0.1 & 1.0 & $>$ 1.4\\
\enddata
Note.
\tablenotetext{a}{
The model employed in this paper 
\citep{Baraffe2015} encompasses a mass range from 0.01 $M_\odot$ to 1.4 $M_\odot$ within the age range of 0.5 to 10 Myr.
Therefore, we write $<$ 0.01 $M_\odot$ or $>$ 1.4 $M_\odot$ for the cases where the estimated detection limit is smaller than the minimum value or larger than the maximum value, respectively.
}
\end{deluxetable*}
\begin{figure*}
\epsscale{1.1}
\gridline{\fig{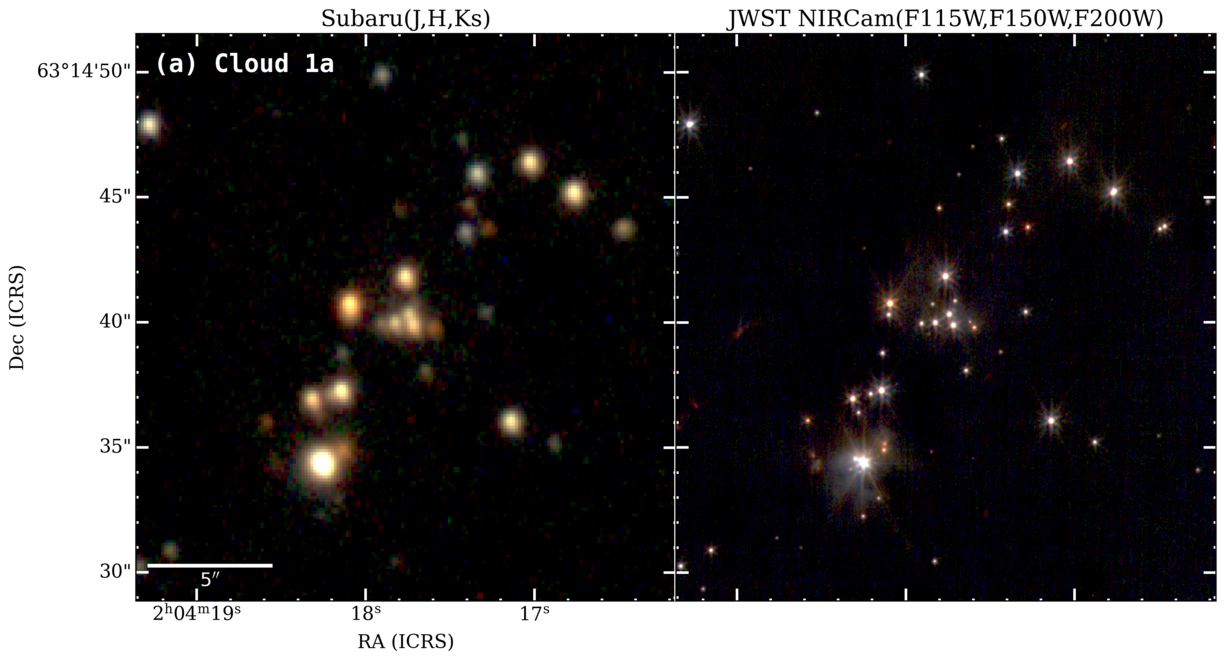}{1.0\textwidth}{}} \vspace{-9mm}
\gridline{\fig{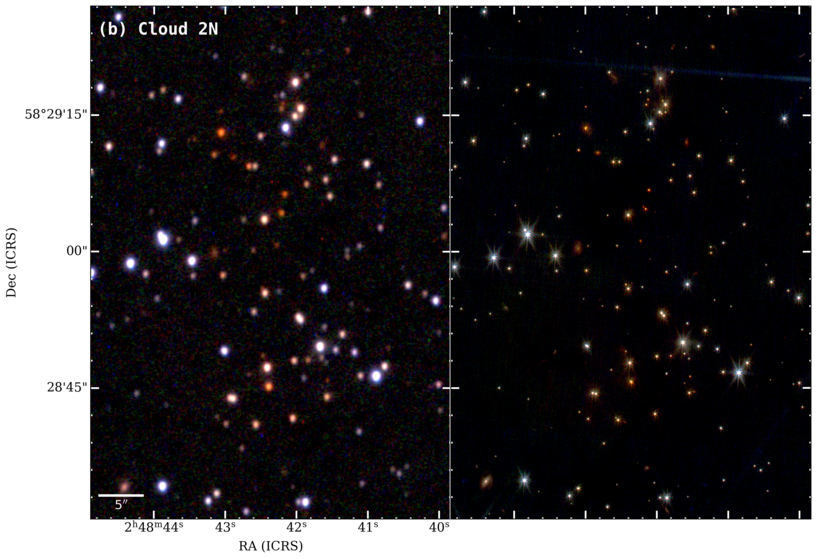}{1.0\textwidth}{} \vspace{-9mm}}
\caption{
IR pseudocolor image of clusters ((a): Cloud 1a, (b): Cloud 2N, (c): Cloud 2S)detected by Subaru (left: $J$, $H$, and $K_S$ bands) and JWST (right: F115W, F150W, and F200W filters).
The displayed area of Cloud 1a and Cloud 2N corresponds to the white dashed square labeled (a) in Figures \ref{entireview_dc1a} and \ref{entireview_dc2n}, respectively.
The displayed area of cloud 2S is within the area of the white dashed square labeled (a) in Figure \ref{entireview_dc2s}.
}
\label{zoom_clusters} 
\end{figure*}
\begin{figure*}
\figurenum{\ref{zoom_clusters}}
\epsscale{1.2}
\plotone{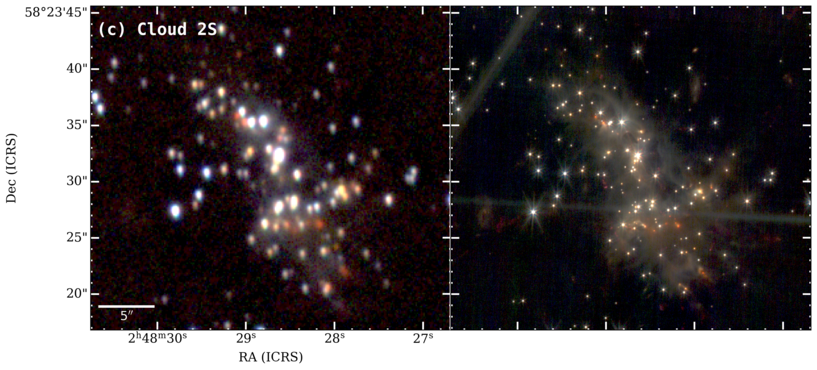}
\caption{(Continued.)
}
\end{figure*}
\begin{figure*}
\epsscale{1.1}
\gridline{\fig{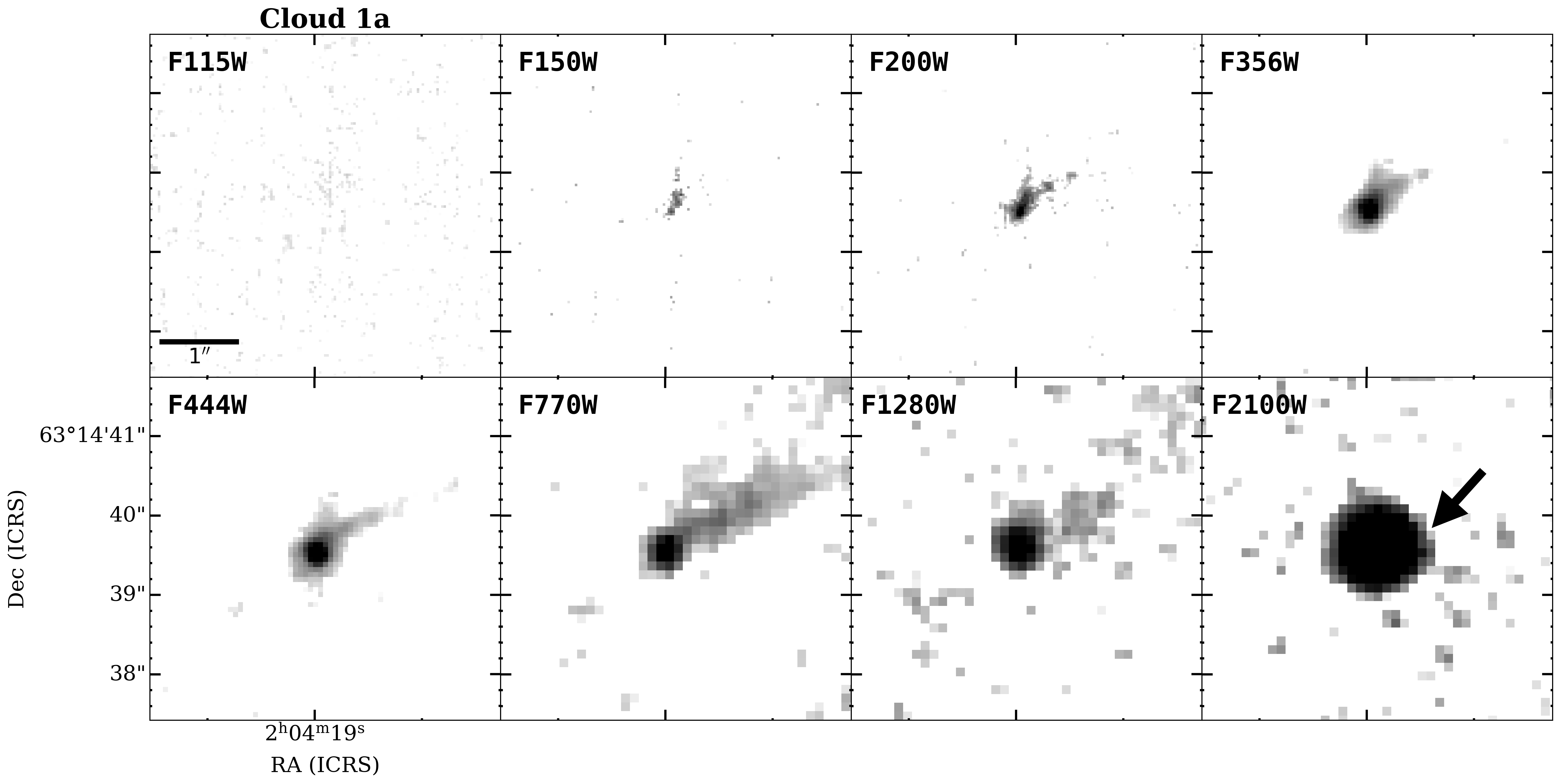}{0.95\textwidth}{}}
\vspace{-0.5cm}
\gridline{\fig{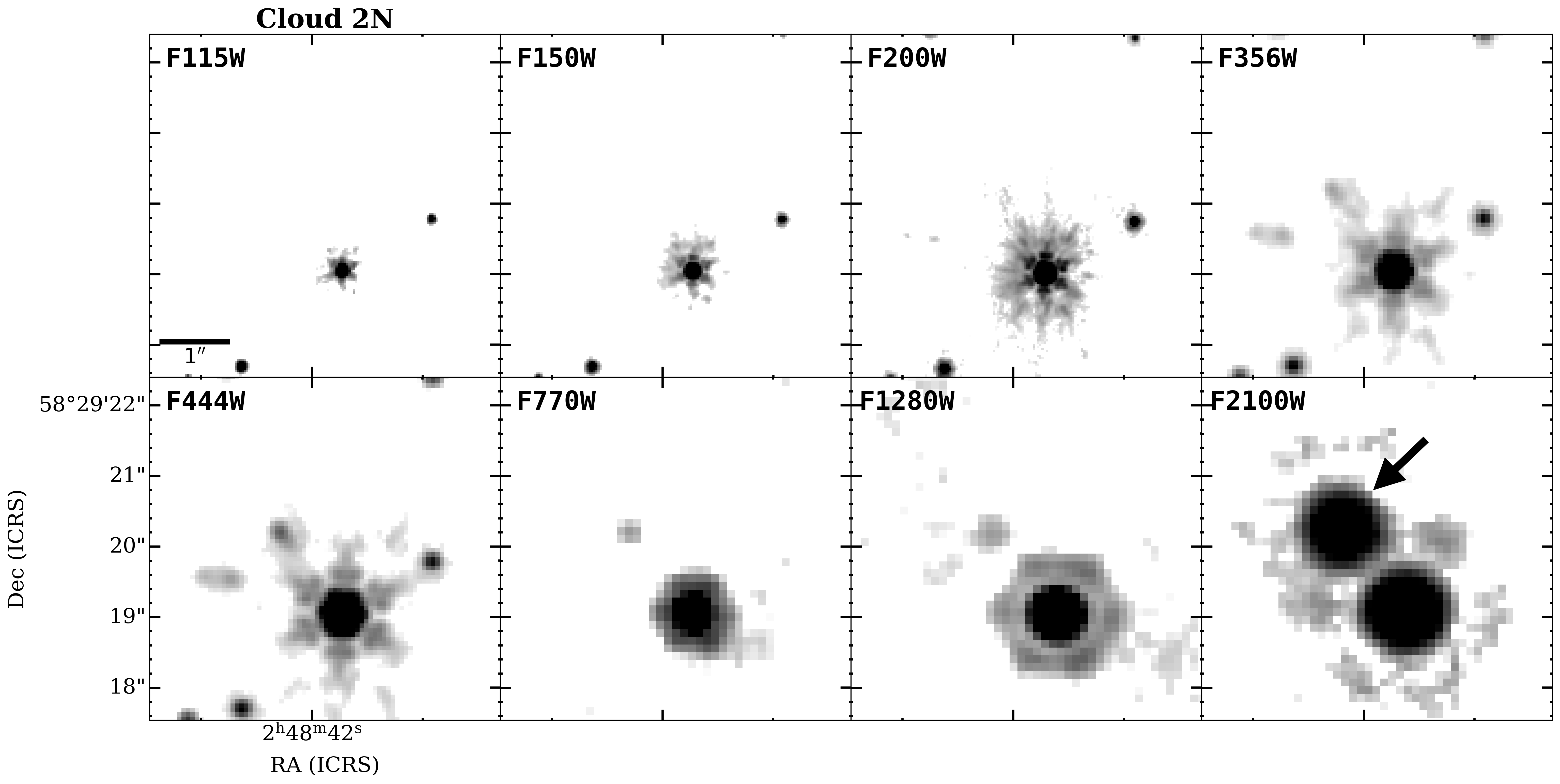}{0.95\textwidth}{}}
\vspace{-0.75cm}
\caption{
NIRCam and MIRI images of candidates of very young sources in Cloud 1a (top) and Cloud 2N (bottom).
Black arrows indicate the positions of the candidates.
The displayed area of Cloud 1a and Cloud 2N corresponds to the white dashed square labeled (c) in Figures \ref{entireview_dc1a} and \ref{entireview_dc2n}, respectively.
}
\label{can_class0} 
\end{figure*}
\begin{figure*}
\gridline{\fig{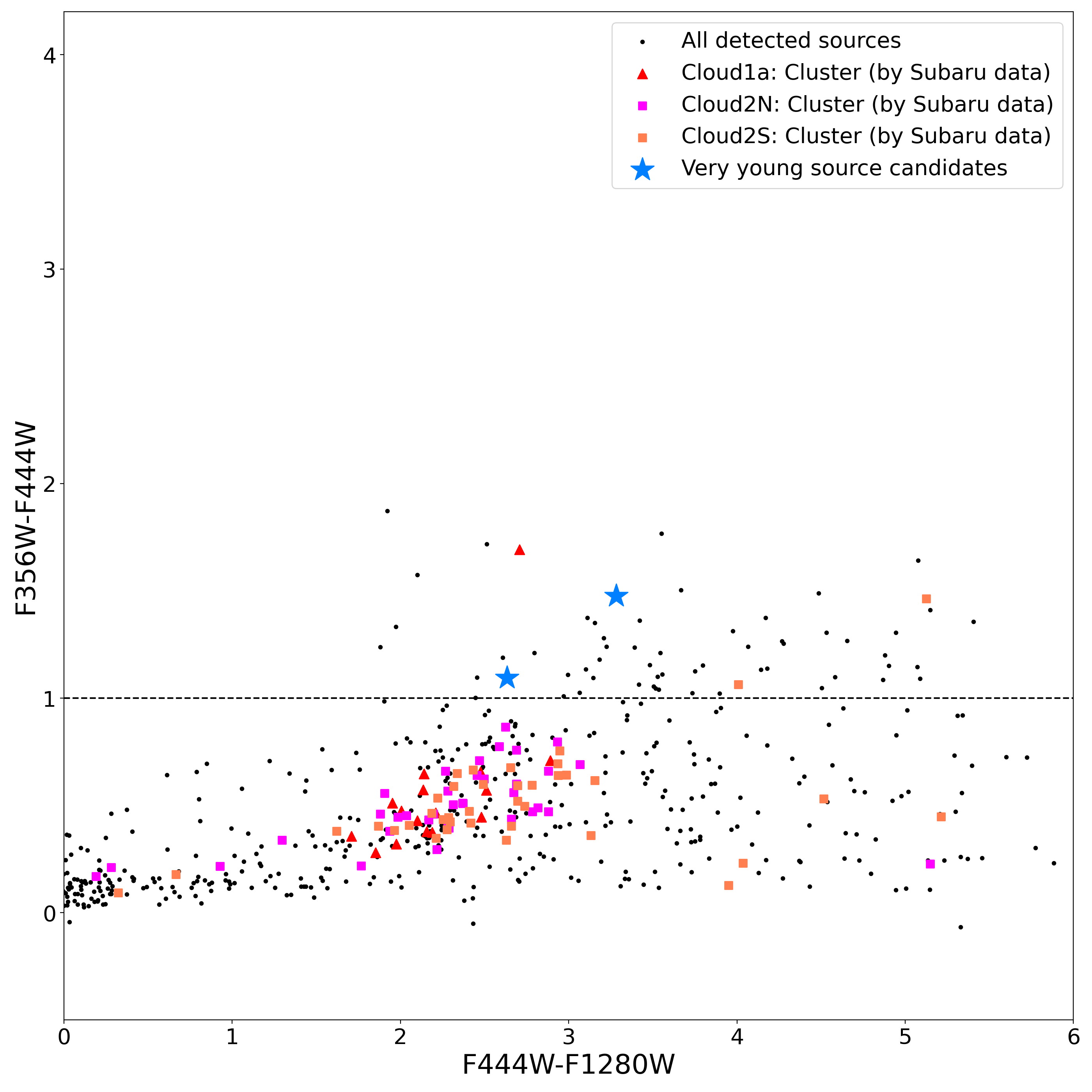}{0.45\textwidth}{}
         \fig{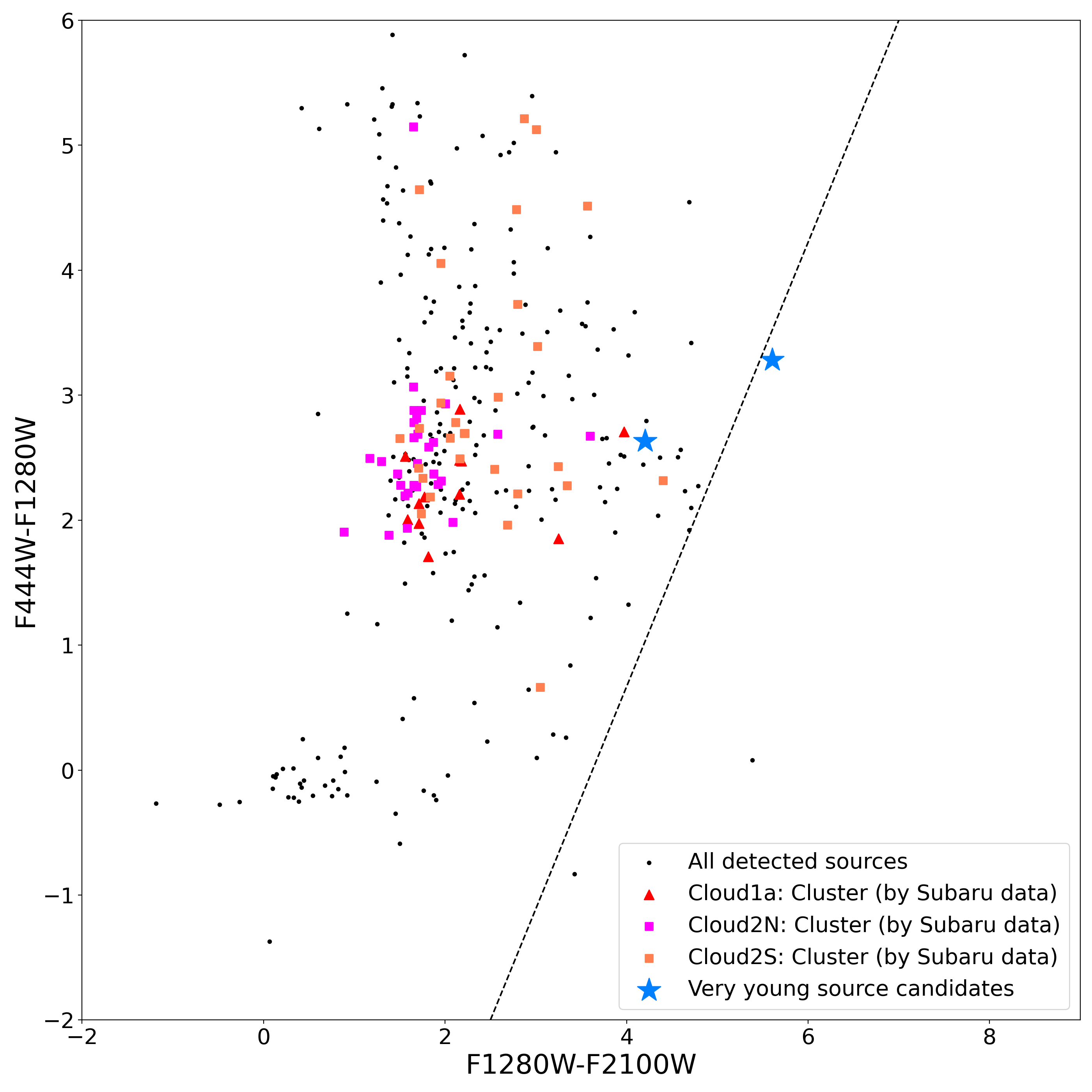}{0.45\textwidth}{}} 
\caption{
Color-color diagrams for all detected sources.
Red triangles, magenta squares, and orange squares indicate the cluster members within Cloud 1a, Cloud 2N, and Cloud 2S, respectively.
These cluster members were identified in Subaru data \citep{Yasui2009, Izumi2014}.
The black dots indicate all sources.
The cyan crosses are the jet/outflow structures.
The blue stars are the very young source candidates.
The black dotted lines indicate the WISE color criteria for class 0 protostars from \citet{Fischer2016}.
}
\label{cc} 
\end{figure*}
\begin{figure*}
\epsscale{1.1}
\gridline{\fig{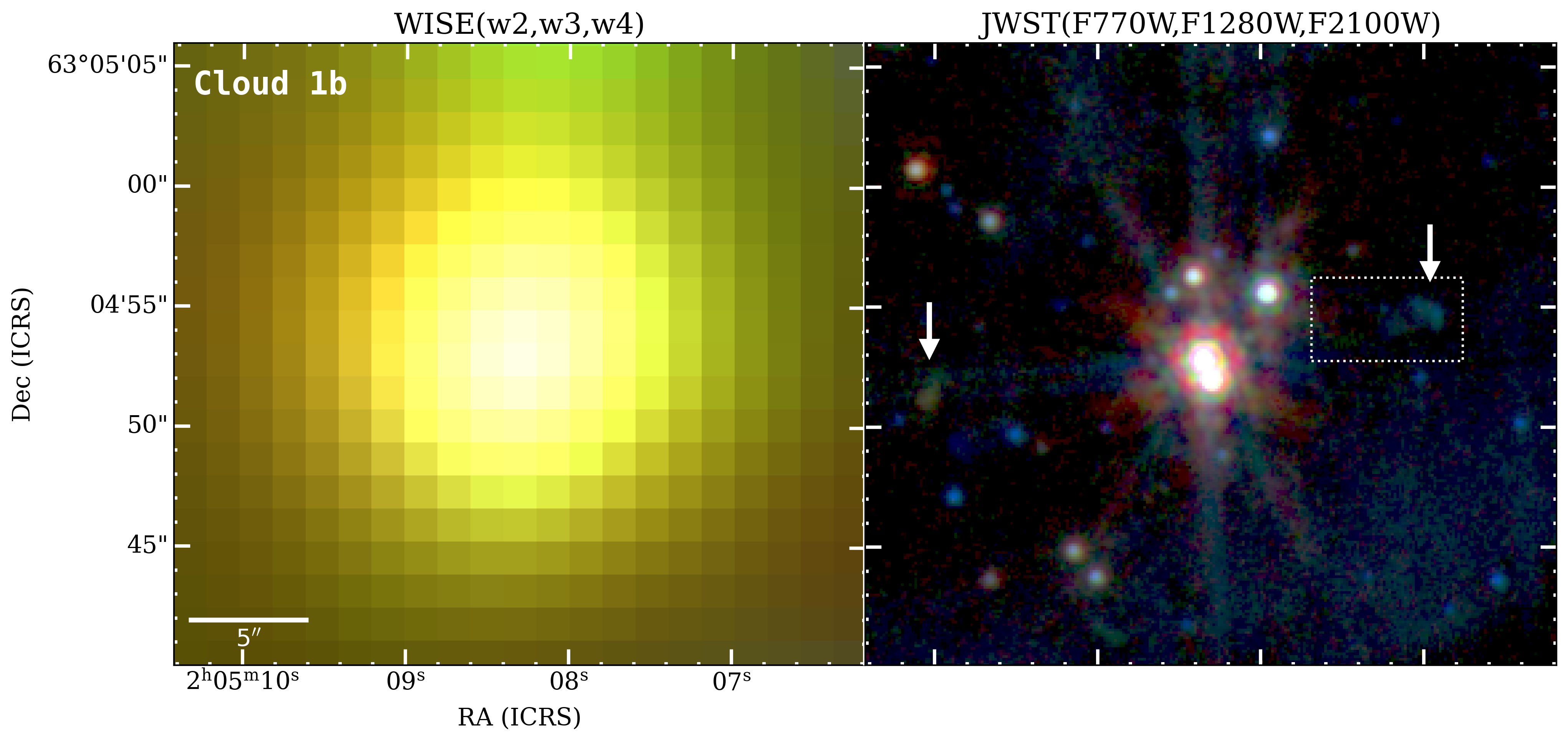}{0.95\textwidth}{}}
\vspace{-0.5cm}
\gridline{\fig{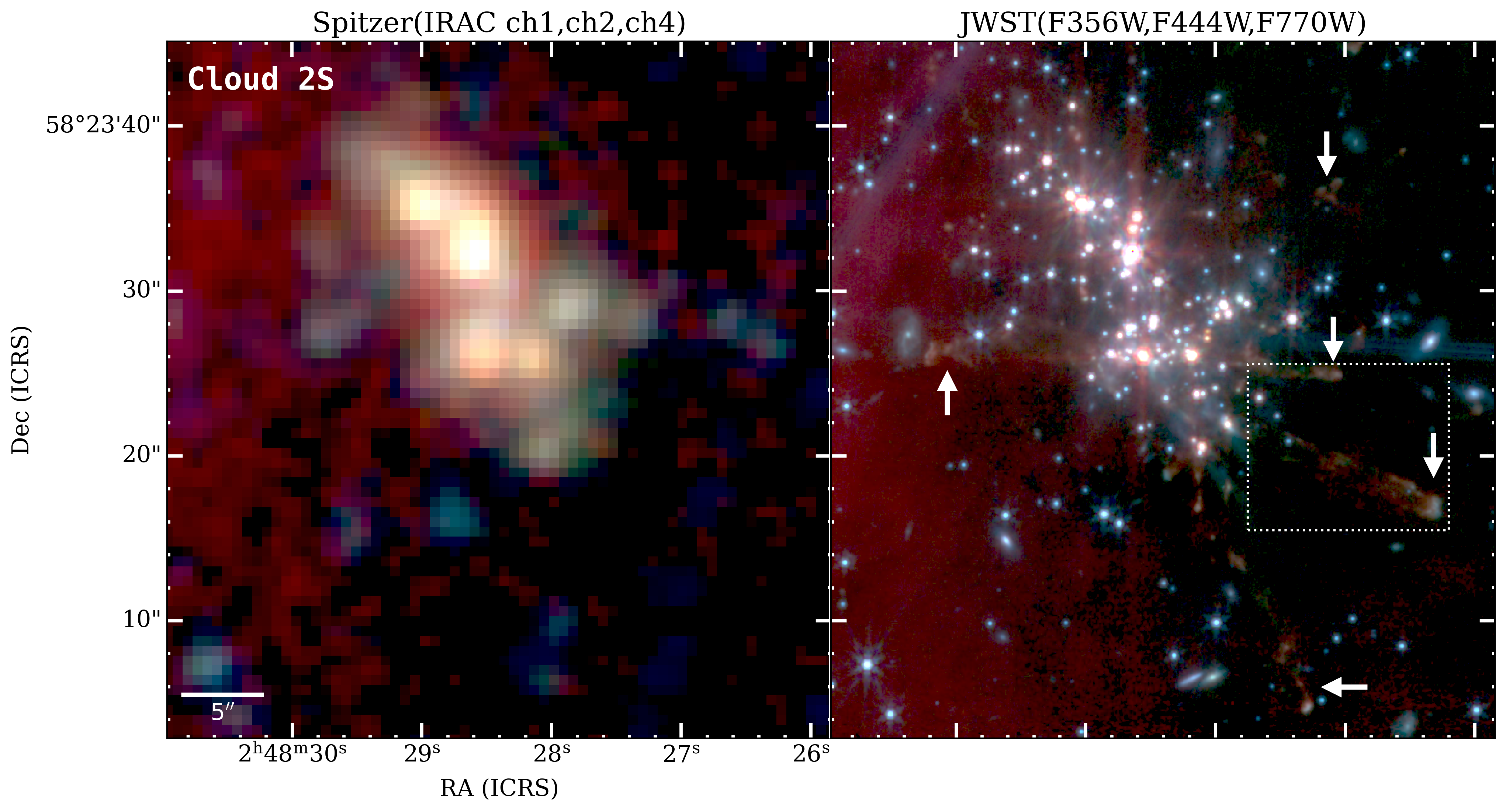}{0.95\textwidth}{}}
\caption{
IR pseudocolor images of possible outflow or jet components within Cloud 1b main cluster (top) and Cloud 2S main cluster (bottom).
The left panels are the images from previous observations (top: WISE with w2, w3, and w4 bands, bottom: Spitzer with IRAC ch1, ch2, and ch4 bands).
The right panels are the new JWST images (top: F770W, F1280W, and F2100W filters; bottom: F356W, F444W, and F770 filters).
The displayed area of Cloud 1b and Cloud 2S corresponds to the white dashed square labeled (b) in Figure \ref{entireview_dc1b} and (a) in Figure \ref{entireview_dc2s}, respectively.
The white arrows indicate the positions of the possible outflow or jet components.
The white dotted rectangles mark the areas shown in Figure \ref{dc1b_detail_outflow} (Cloud 1b) and \ref{dc2s_detail_outflow} (Cloud 2N).
}
\label{zoom_outflows} 
\end{figure*}
\begin{figure*}
\epsscale{0.7}
\plotone{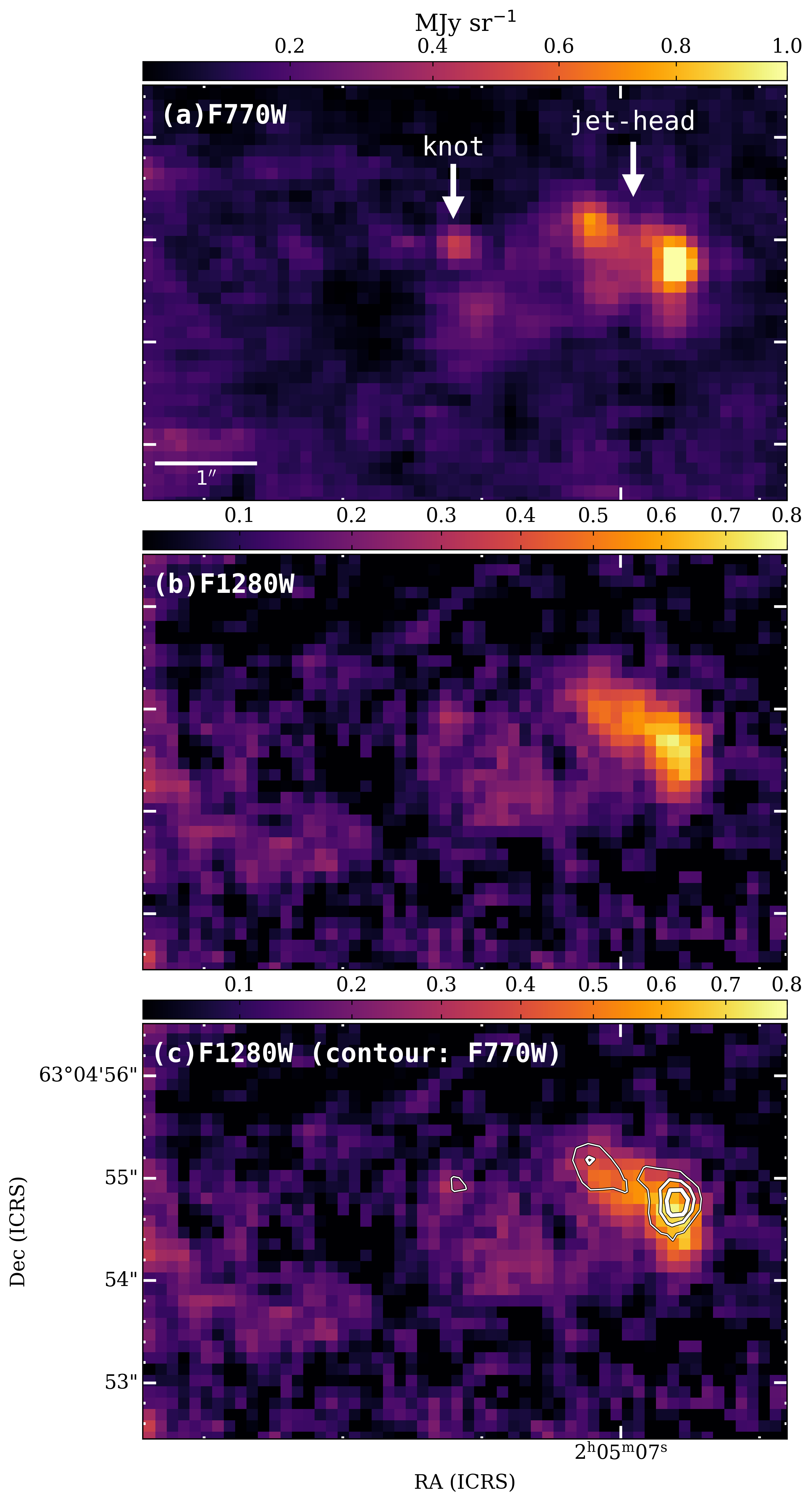}
\caption{
MIRI images with F770W (a) and F1280W (b) filters for possible outflow or jet components within the Cloud 1b main cluster (white dotted rectangle in the top right panel of Figure \ref{zoom_outflows}).
The panel (c) shows the comparison between F770W (white contours) and F1280W (background) filters.
The contours indicate 3$\sigma$, 5$\sigma$, and 7$\sigma$ where 1$\sigma$ is 0.15 MJy sr$^{-1}$.
The width of the contours increases with the contour level.
}
\label{dc1b_detail_outflow} 
\end{figure*}
\begin{figure*}
\epsscale{1.0}
\plotone{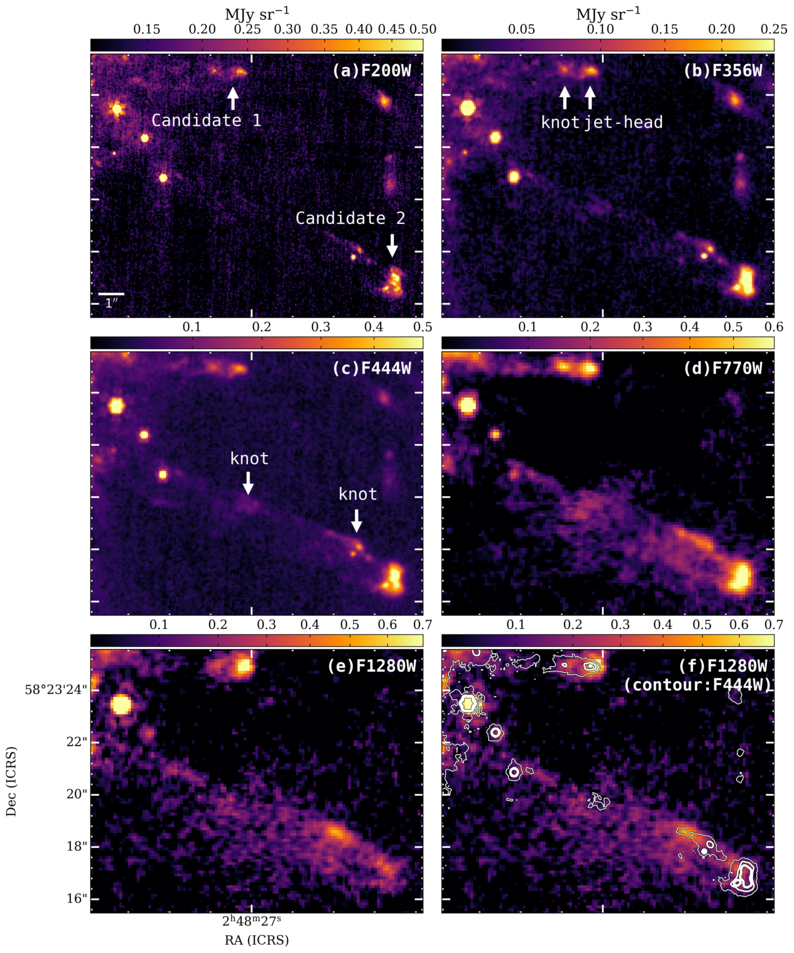}
\caption{
NIRCam and MIRI images with F200W (a), F356W (b), F444W (c), F770W (d), and F1280W (e)
for two representative possible outflow or jet components within the Cloud 2S main cluster (white dotted rectangle in the bottom right panel of Figure \ref{zoom_outflows}).
Panel (f) shows the comparison between F444W (white contours) and F1280W (background) filters.
The contours indicate 3$\sigma$, 9$\sigma$, and 15$\sigma$ where 1$\sigma$ is 0.025 MJy sr$^{-1}$.
The width of the contours increases with the contour level.
}
\label{dc2s_detail_outflow} 
\end{figure*}
\begin{figure*}
\epsscale{1.1}
\gridline{\fig{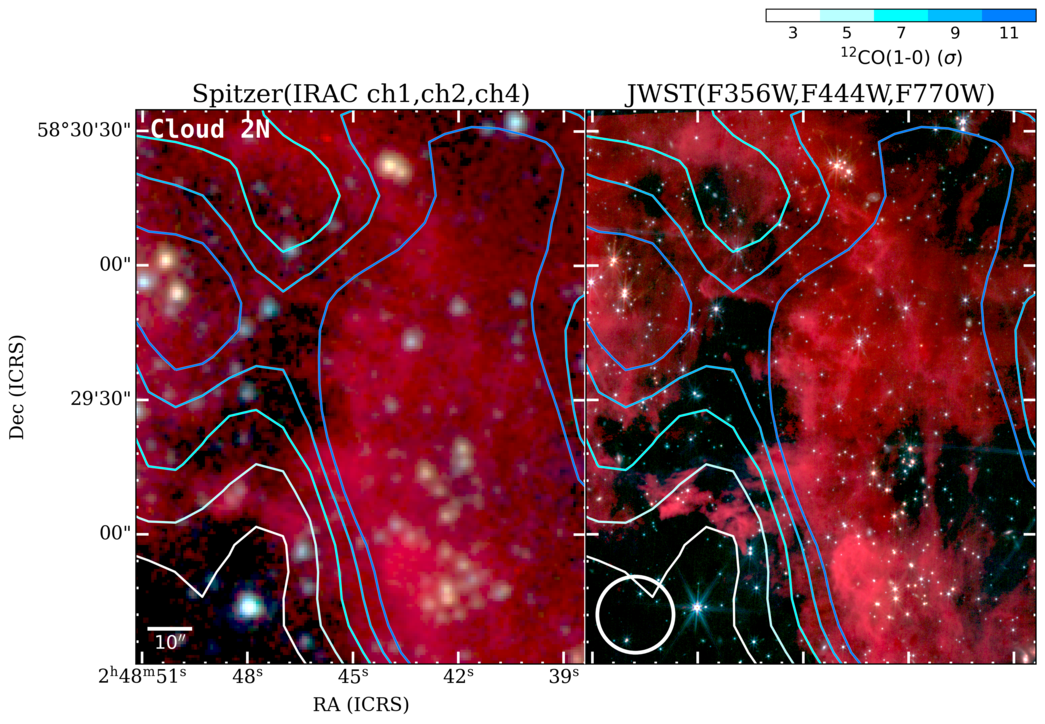}{0.95\textwidth}{}}
\vspace{-0.5cm}
\gridline{\fig{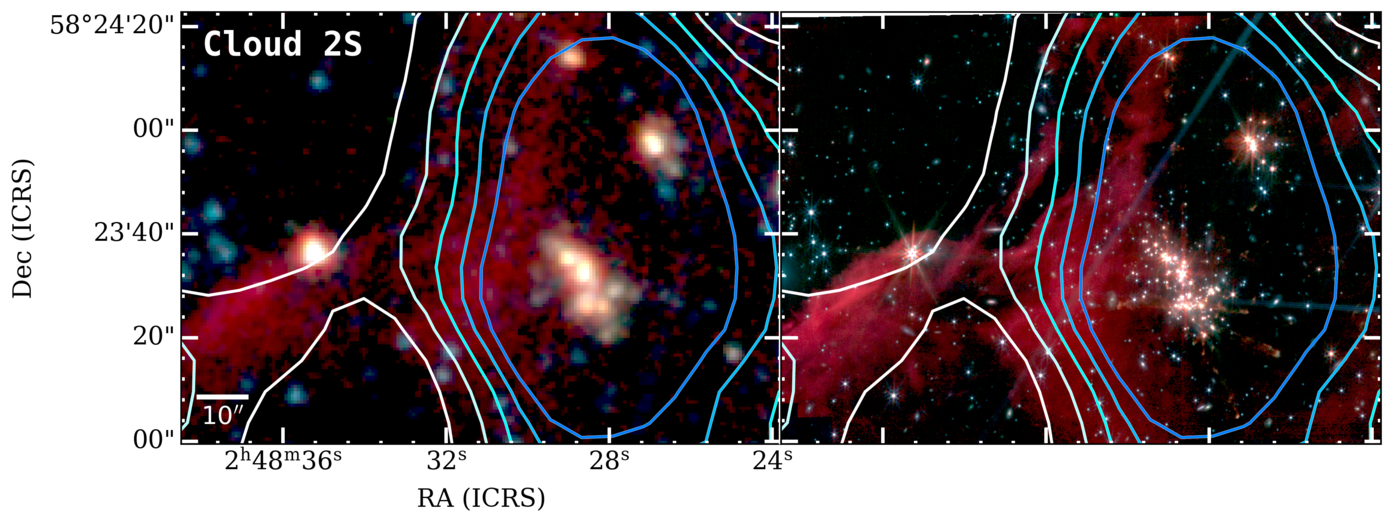}{0.95\textwidth}{}}
\caption{
IR pseudocolor images of nebular structures in Cloud 2N (top) and 2S (bottom).
The left panels are the images from previous Spitzer observations (IRAC ch1, ch2, and ch4 bands).
The right panels are the new JWST images  (F356W, F444W, and F770 filters).
The contours show the integrated $^{12}$CO(1-0) map of Digel Cloud 2 from our NRO 45m data
\citep[$V_{\rm LSR}$ = -106.1 -- -99.1 km s$^{-1}$;][]{Izumi2017}, with contour levels at 3$\sigma$, 5$\sigma$, 7$\sigma$, 9$\sigma$, and 11$\sigma$, where  1$\sigma$ = 1.35 K km s$^{-1}$
(same as in the bottom right panel in Figure \ref{FoV}).
The color bar located in the top right indicates the contour level.
The white circle at the lower light corner in the top right panel shows the beam size of the NRO 45m ($\sim$17$^{\prime \prime}$).
}
\label{nebula} 
\end{figure*}
\clearpage
\appendix
\section{Observation Settings}\label{sec:a-1}
In this appendix, we report the detailed information of the observation settings.
Table \ref{obs-para} summarizes the parameters of all observations.

\section{all data}\label{sec:a-2}
\begin{figure*}
\epsscale{1.1}
\plotone{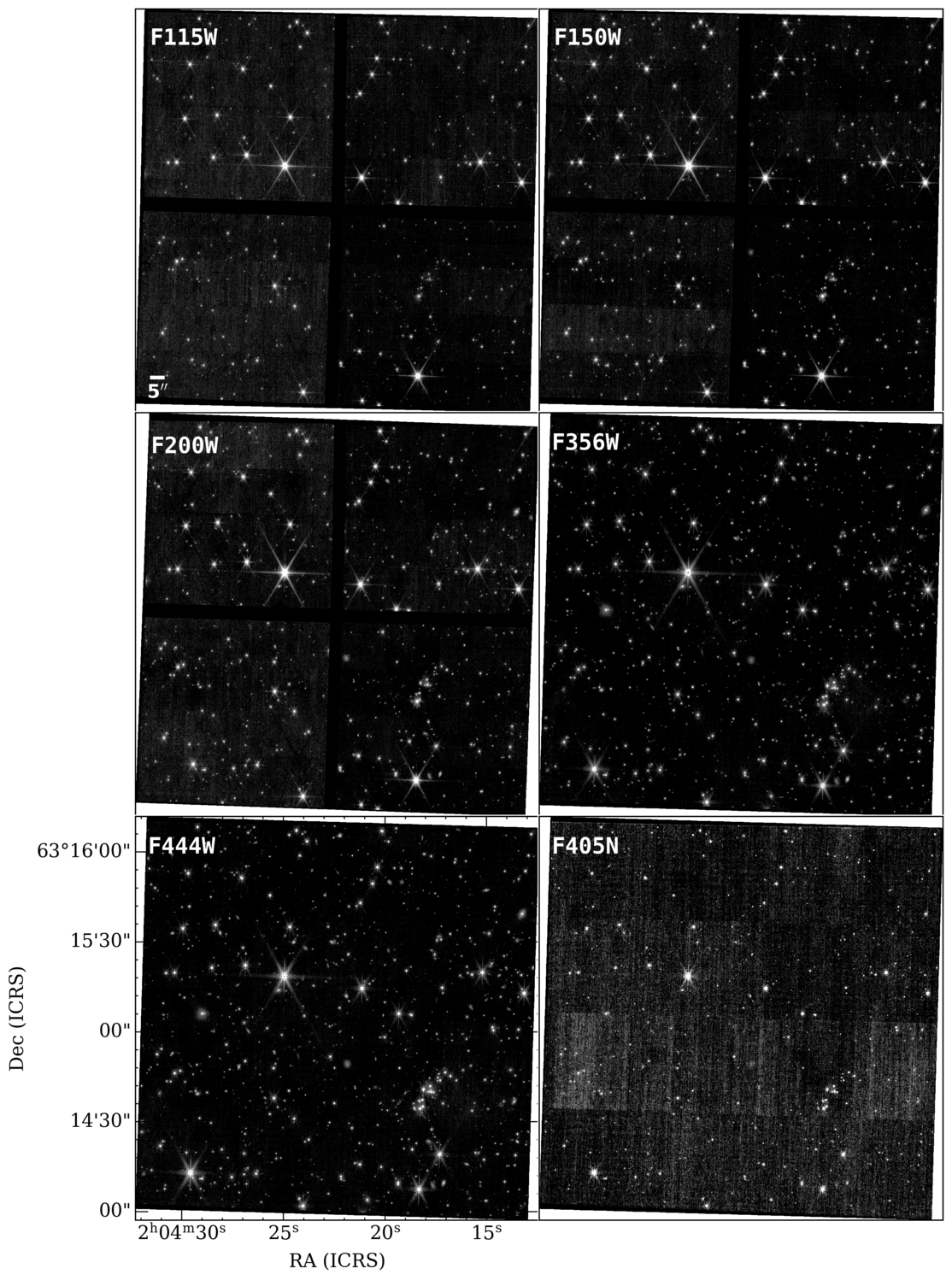}
\caption{All NIRCam monochromatic images of DC 1a.}
\label{dc1a_nircam}
\end{figure*}

\begin{figure*}
\epsscale{1.1}
\plotone{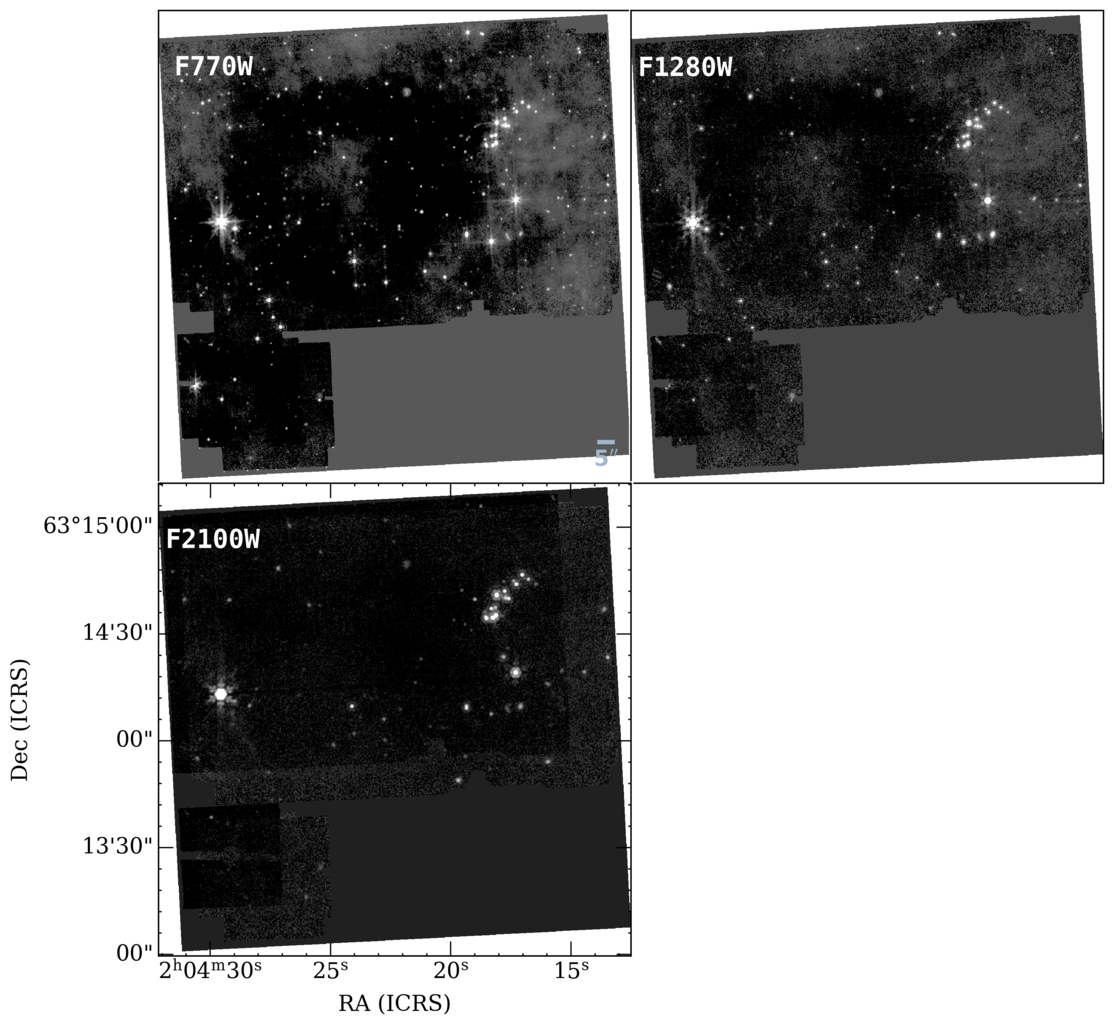}
\caption{All MIRI monochromatic images of DC 1a.}
\label{dc1a_miri}
\end{figure*}

\begin{figure*}
\epsscale{1.1}
\plotone{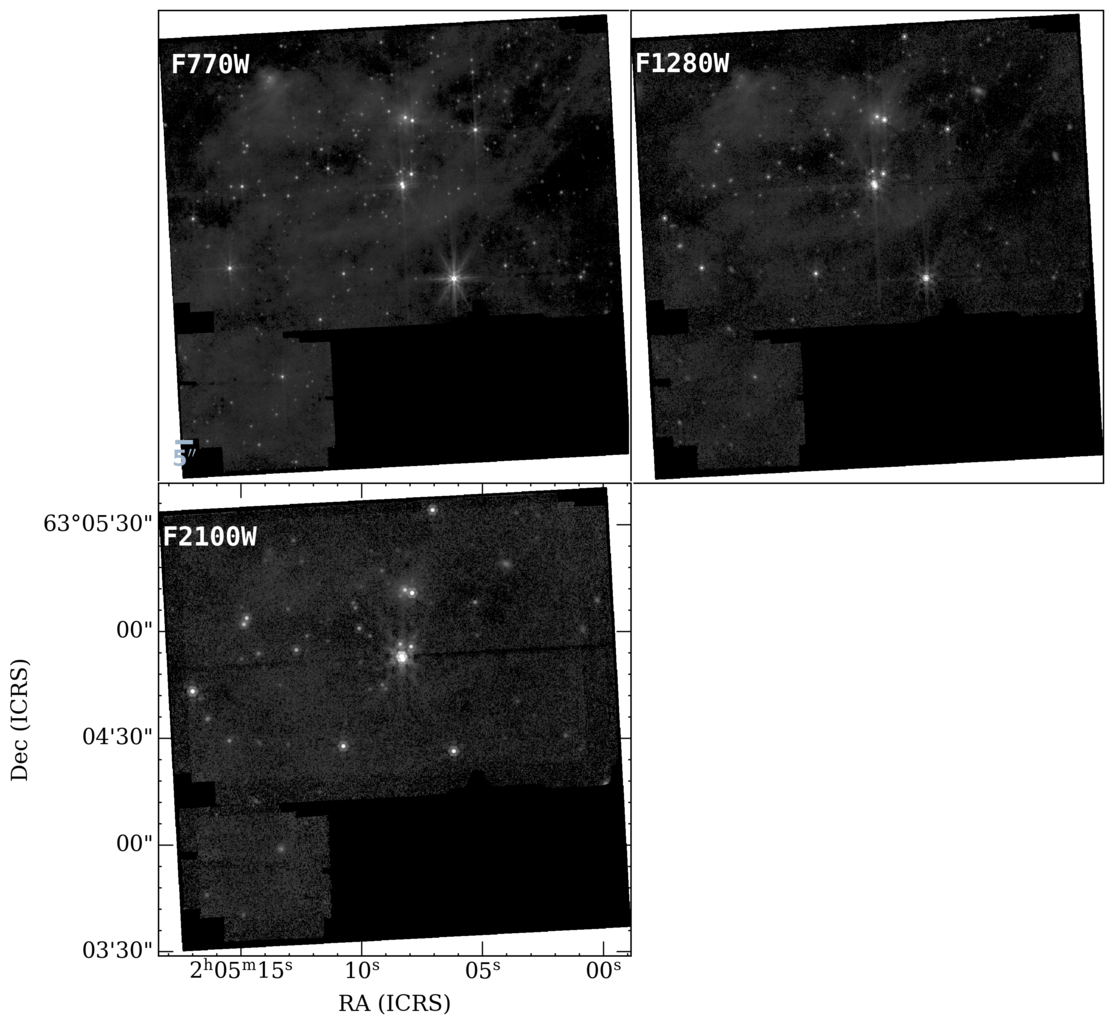}
\caption{All MIRI monochromatic images of DC 1b.
}
\label{dc1b_miri}
\end{figure*}

\begin{figure*}
\epsscale{1.1}
\plotone{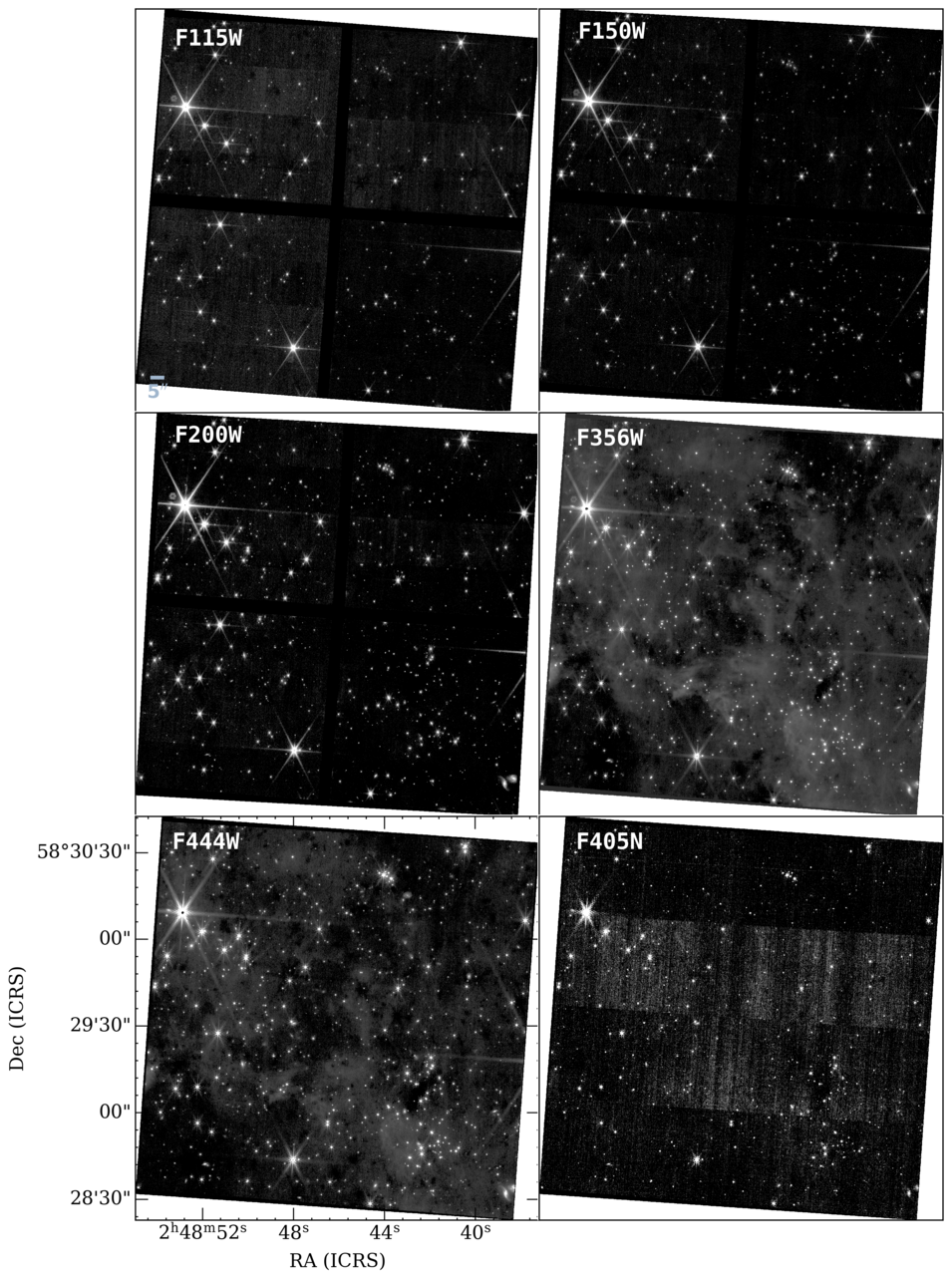}
\caption{All NIRCam monochromatic images of DC 2N.
}
\label{dc2n_nircam}
\end{figure*}

\begin{figure*}
\epsscale{1.1}
\plotone{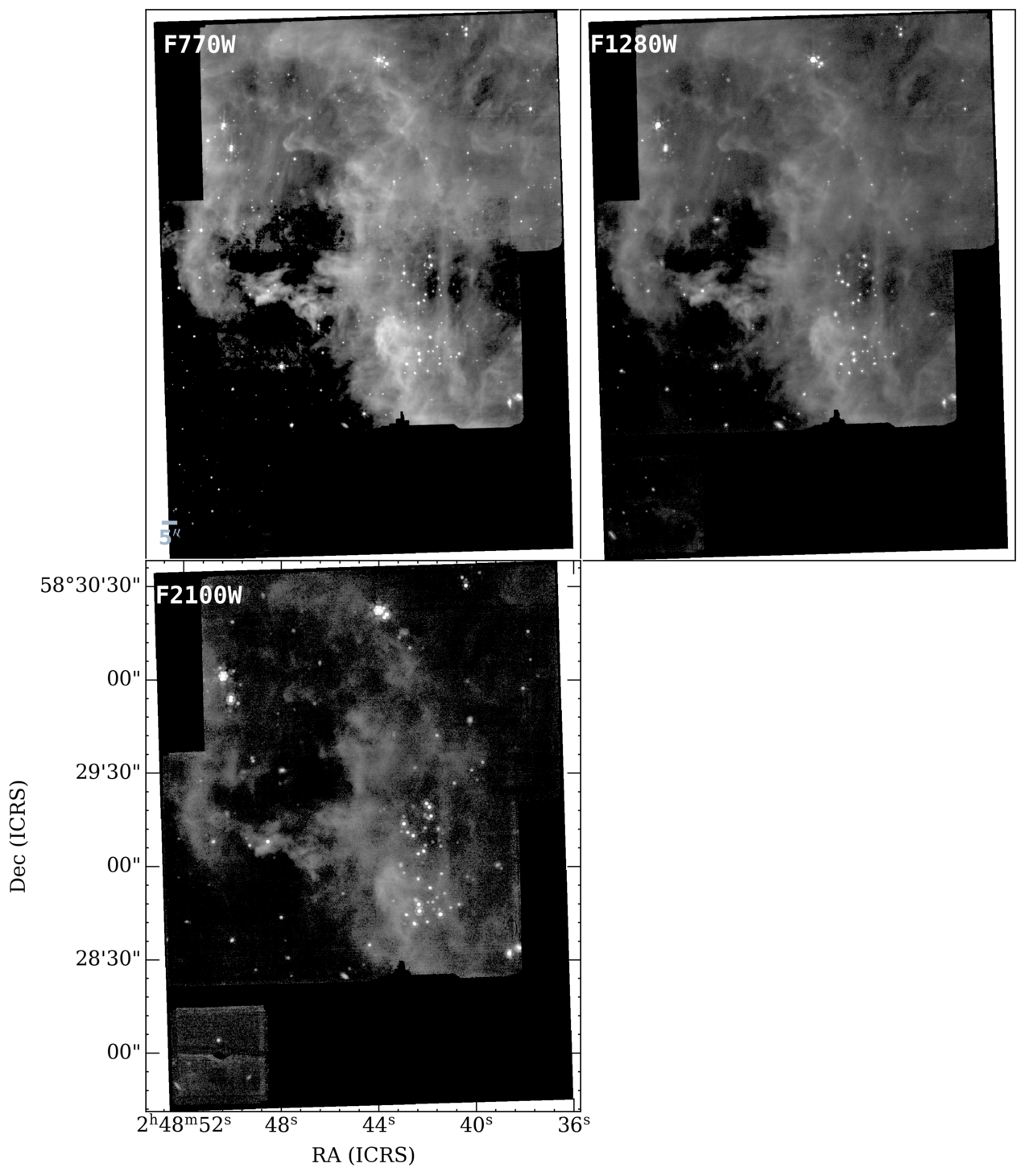}
\caption{All MIRI monochromatic images of DC 2N.
}
\label{dc2n_miri}
\end{figure*}

All images obtained through this project (PID1237) are shown in this appendix.
Figures \ref{dc1a_nircam} and \ref{dc1a_miri} show all the NIRCam and MIRI monochromatic images of Cloud 1a.
Figure \ref{dc1b_miri} displays all the MIRI monochromatic images of Cloud 1b.
Figures \ref{dc2n_nircam} and \ref{dc2n_miri} present all the NIRCam and MIRI monochromatic images of Cloud 2N.
Figures \ref{dc2s_nircam} and \ref{dc2s_miri} show all the NIRCam and MIRI monochromatic images of Cloud 2S.
The horizontal stripes in some of the NIRCam images, especially those taken with the lower-signal F405 filter are due to 1/f noise that is not yet mitigated in the released pipeline \citep{Bagley2023}.
\begin{longrotatetable}
\begin{deluxetable*}{ccccccccccccc}
\tablecaption{Parameters of JWST observations. \label{obs-para}}
\tablewidth{0pt}
\tablehead{
\colhead{Target} & \colhead{} & \colhead{Module} & \colhead{Subarray} &
\colhead{Filter} & \colhead{Dither} & \colhead{Mosaic} & \colhead{Readout} &
\colhead{Groups} & \colhead{Integration} & \colhead{Exposures} &
\colhead{Total Exposure}  \\ 
                  &                     &                   &              &
                  & \colhead{pattern} & \colhead{number} & \colhead{pattern} &
\colhead{/Int} & \colhead{/Exp} &\colhead{/Dith} & 
\colhead{time}  \\ 
}
\startdata
Cloud 1a & NIRCam  & B & FULL & F115W\tablenotemark{a}  & Subpixel STANDARD 4 & 1 & RAPID & 8 & 1 & --- & 343.577 \\ 
         &         & B &  FULL & F150W\tablenotemark{a}  & Subpixel STANDARD 4 & 1 & RAPID & 8 & 1 & --- & 343.577\\ 
         &         & B &  FULL & F200W\tablenotemark{a}  & Subpixel STANDARD 4 & 1 & RAPID & 8 & 1 & --- & 343.577\\ 
         &         & B &  FULL & F356W\tablenotemark{a}  & Subpixel STANDARD 4 & 1 & RAPID & 8 & 1 & --- & 343.577\\ 
         &         & B &  FULL & F444W\tablenotemark{a}  & Subpixel STANDARD 4 & 1 & RAPID & 8 & 1 & --- & 343.577\\ 
         &         & B &  FULL & F405N\tablenotemark{a}  & Subpixel STANDARD 4 & 1 & RAPID & 8 & 1 & --- & 343.577\\ 
         & MIRI    & --- & FULL & F770W  & CYCLING 4 & 1 & FASTR1 & 36 & 1 & 1 & 399.606 \\ 
         &         & --- & FULL & F1280W & CYCLING 4 & 1 & FASTR1 & 36 & 1 & 1 & 399.606\\ 
         &         & --- & FULL & F2100W & CYCLING 4 & 1 & FASTR1 & 36 & 1 & 1 & 399.606\\  \hline
Cloud 1b & MIRI    & --- & FULL & F770W  & CYCLING 4 & 1 & FASTR1 & 36 & 1 & 1 & 399.606 \\ 
         &         & --- & FULL & F1280W & CYCLING 4 & 1 & FASTR1 & 36 & 1 & 1 & 399.606\\ 
         &         & --- & FULL & F2100W & CYCLING 4 & 1 & FASTR1 & 36 & 1 & 1 & 399.606\\  \hline
Cloud 2N & NIRCam  & B &  FULL & F115W  & Subpixel STANDARD 4 & 1 & RAPID & 8 & 1 & --- & 343.577 \\ 
         &         & B &  FULL & F150W  & Subpixel STANDARD 4 & 1 & RAPID & 8 & 1 & --- & 343.577\\ 
         &         & B &  FULL & F200W  & Subpixel STANDARD 4 & 1 & RAPID & 8 & 1 & --- & 343.577\\ 
         &         & B &  FULL & F356W  & Subpixel STANDARD 4 & 1 & RAPID & 8 & 1 & --- & 343.577\\ 
         &         & B &  FULL & F444W  & Subpixel STANDARD 4 & 1 & RAPID & 8 & 1 & --- & 343.577\\ 
         &         & B &  FULL & F405N  & Subpixel STANDARD 4 & 1 & RAPID & 8 & 1 & --- & 343.577\\ 
         & MIRI\tablenotemark{b}    &--- &  FULL & F770W  & CYCLING 4 & 2 &  FASTR1 & 35 & 1 & 1 & 388.506 \\ 
         &         & --- & FULL & F1280W & CYCLING 4 & 2 &  FASTR1 & 35 & 1 & 1 & 388.506\\ 
         &         & --- & FULL & F2100W & CYCLING 4 & 2 &  FASTR1 & 35 & 1 & 1 & 388.506\\  \hline
Cloud 2S & NIRCam\tablenotemark{b}  & B &  FULL & F115W  & Subpixel STANDARD 4 & 2 & BRIGHT2 & 5 & 1 & --- & 429.471 \\ 
         &         & B &  FULL & F150W  & Subpixel STANDARD 4 & 2 & BRIGHT2 & 5 & 1 & --- & 429.471\\ 
         &         & B &  FULL & F200W  & Subpixel STANDARD 4 & 2 & BRIGHT2 & 5 & 1 & --- & 429.471\\ 
         &         & B &  FULL & F356W  & Subpixel STANDARD 4 & 2 & BRIGHT2 & 5 & 1 & --- & 429.471\\ 
         &         & B &  FULL & F444W  & Subpixel STANDARD 4 & 2 & BRIGHT2 & 5 & 1 & --- & 429.471\\ 
         &         & B &  FULL & F405N  & Subpixel STANDARD 4 & 2 & BRIGHT2 & 5 & 1 & --- & 429.471\\ 
         & MIRI    & --- & FULL & F770W  & CYCLING 4 & 1 & FASTR1 & 36 & 1 & 1 & 399.606 \\ 
         &         & --- & FULL & F1280W & CYCLING 4 & 1 & FASTR1 & 36 & 1 & 1 & 399.606\\ 
         &         & --- & FULL & F2100W & CYCLING 4 & 1 & FASTR1 & 37 & 1 & 1 & 410.706\\  \hline
\enddata
\tablenotetext{a}{All NIRCam imaging, short-filter (F115W, F150W, F200W) and long-filter (F356W, F444W, F405N+F444W) observations were performed simultaneously.}
\tablenotetext{b}{NIRCam observations of Cloud 2S and MIRI observations of Cloud 2N were performed simultaneously based on the NIRCam-MIRI parallel observations.}
\end{deluxetable*}
\end{longrotatetable}
\begin{figure*}
\epsscale{1.1}
\plotone{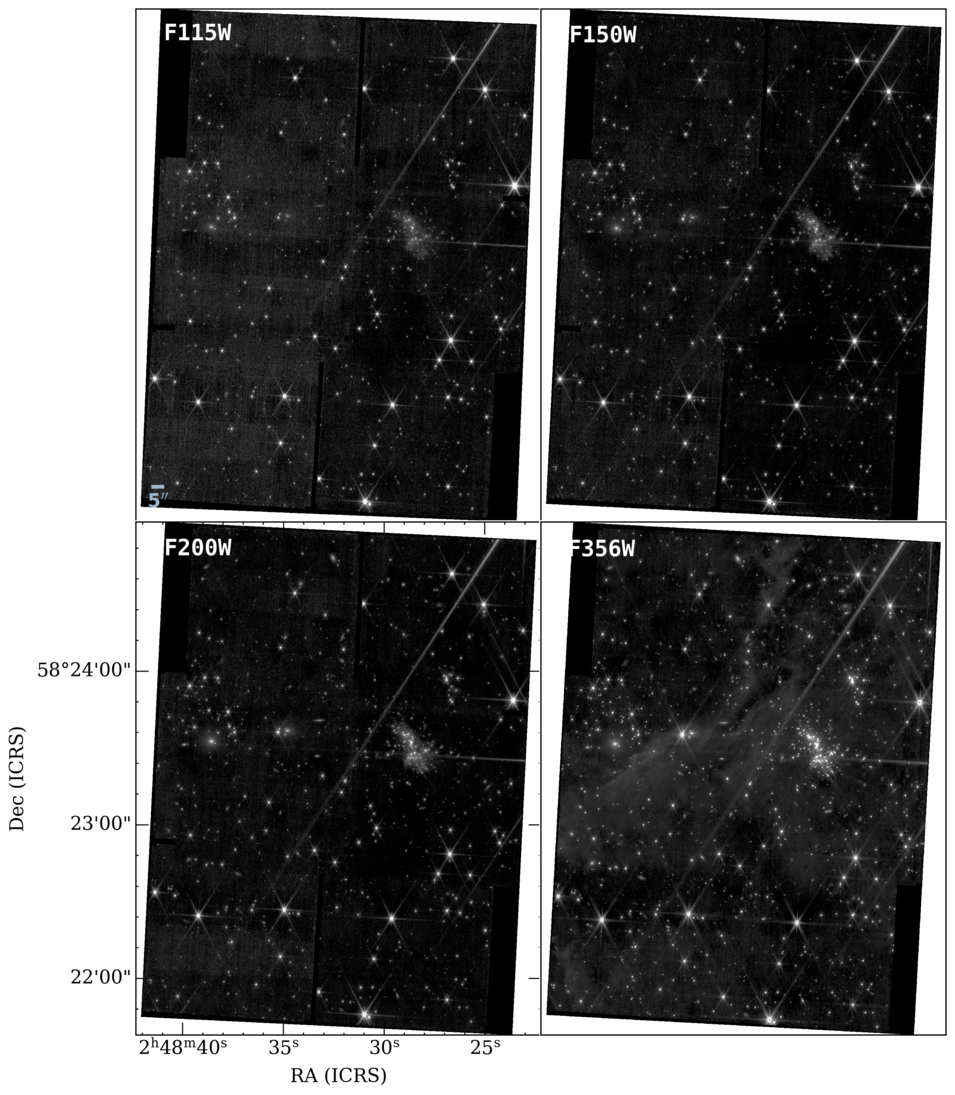}
\caption{All NIRCam monochromatic images of DC 2S.
}
\label{dc2s_nircam}
\end{figure*}
\begin{figure*}
\figurenum{\ref{dc2s_nircam}}
\epsscale{1.1}
\plotone{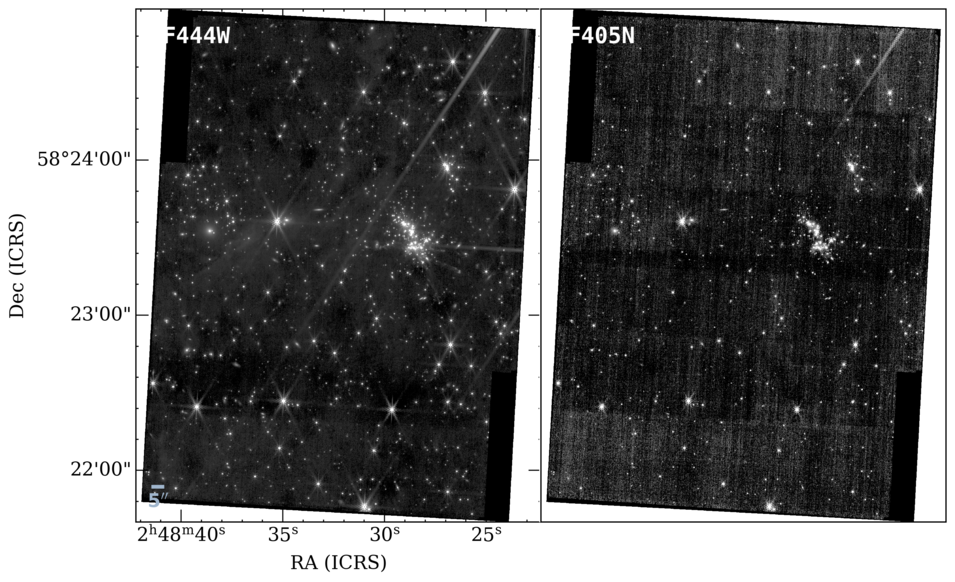}
\caption{(Continued.)
}
\end{figure*}
\begin{figure*}
\epsscale{1.1}
\plotone{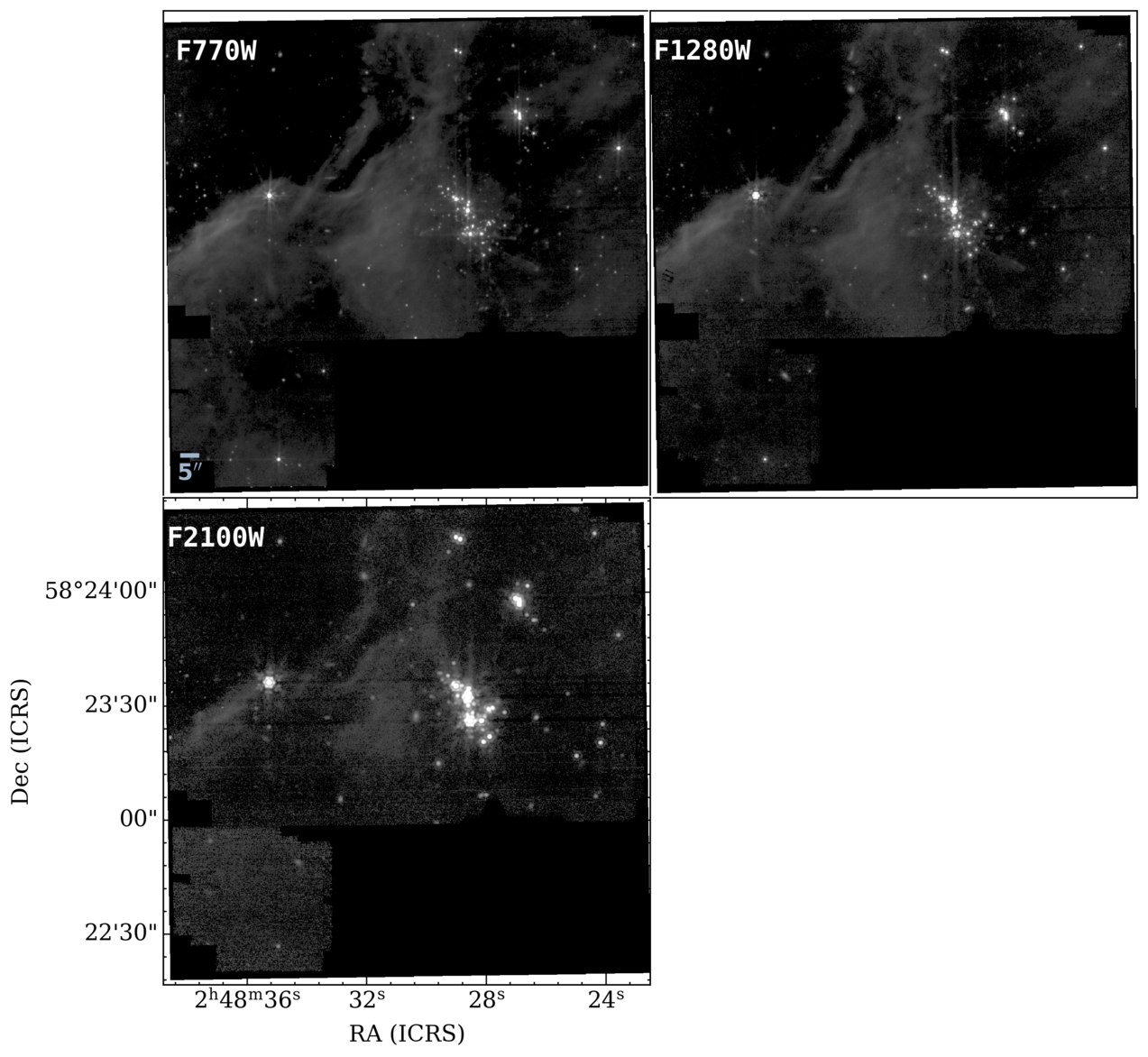}
\caption{All MIRI monochromatic images of DC 2S.
}
\label{dc2s_miri}
\end{figure*}



\bibliographystyle{aasjournal}
\bibliography{Ref}


\end{document}